\documentclass{aa}
\usepackage{graphicx}
\usepackage{txfonts}
\usepackage[dvipsnames]{xcolor}
\usepackage{marvosym}
\usepackage{natbib}
\bibpunct{(}{)}{;}{a}{}{,}
\usepackage[none]{hyphenat} 
\usepackage{mathtools}
\usepackage{amsmath}
\usepackage{soul}
\usepackage{subcaption,booktabs}
\usepackage{multirow}
\usepackage{comment}
\usepackage{lscape}
\usepackage[thinc]{esdiff}

\definecolor{darkblue}{RGB}{12,94,176}
\usepackage{hyperref}
\hypersetup{hidelinks}
\hypersetup{
    pdfnewwindow=true,     
    colorlinks=true,       
    linkcolor=darkblue,    
    citecolor=darkblue,    
    filecolor=darkblue,    
    urlcolor=blue          
}

\usepackage{siunitx}

\DeclareSIUnit\parsec{pc}
\DeclareSIUnit\cMpc{cMpc}
\DeclareSIUnit\year{yr}
\DeclareSIUnit\Zsun{Z_{\odot}}
\DeclareSIUnit\Msun{M_{\odot}}
\DeclareSIUnit\Rsun{R_{\odot}}
\DeclareSIUnit\Lsun{L_{\odot}}
\DeclareSIUnit\erg{erg}
\DeclareSIUnit\eV{eV}

\begin{document}

   \title{Milky Way globular clusters: Nurseries for dynamically formed binary black holes}

   \author{
   Federico Angeloni\inst{1,2,3,4} \thanks{\email{federico.angeloni@students.uniroma2.eu}}
   \and
   Konstantinos Kritos\inst{5}
   \and
   Raffaella Schneider\inst{1,3,4}
   \and
   Emanuele Berti\inst{5}
   \and 
   Luca Graziani\inst{1,3,4}
   \and
   Stefano Torniamenti\inst{6, 7}
   \and
   Michela Mapelli\inst{7, 8, 9, 10, 11}
   }

   \institute{
   Dipartimento di Fisica, Sapienza, Università di Roma, Piazzale Aldo Moro 5, 00185, Roma, Italy\and
   Dipartimento di Fisica, Tor Vergata, Università di Roma, via della Ricerca Scientifica, 00133, Roma, Italy\and
   INAF/Osservatorio Astronomico di Roma, Via di Frascati 33, 00078 Monte Porzio Catone, Italy\and
   INFN, Sezione di Roma I, Piazzale Aldo Moro 2, 00185 Roma, Italy\and
   William H. Miller III Department of Physics and Astronomy, Johns Hopkins University, 3400 N. Charles Street, Baltimore, Maryland, 21218, USA\and
   Max-Planck-Institut für Astronomie, Königstuhl 17, 69117, Heidelberg, Germany\and
   Universität Heidelberg, Zentrum für Astronomie (ZAH), Institut für Theoretische Astrophysik, Albert Ueberle Str. 2, 69120, Heidelberg, Germany\and
   Universität Heidelberg, Interdisziplinäres Zentrum für Wissenschaftliches Rechnen, Heidelberg, Germany\and
   Physics and Astronomy Department Galileo Galilei, University of Padova, Vicolo dell’Osservatorio 3, 35122 Padova, Italy\and
   INFN – Padova, Via Marzolo 8, 35131 Padova, Italy\and
   INAF – Osservatorio Astronomico di Padova, Vicolo dell’Osservatorio 5, 35122 Padova, Italy
   }

   \date{Received ...; accepted ...}

  \abstract
  % {} leave it empty if necessary  
   {We present a novel self-consistent theoretical framework to characterize the formation, evolution, and merger sites of dynamically formed black hole binaries, with a focus on explaining the most massive events observed by the LIGO-Virgo-KAGRA Collaboration. Our approach couples the galaxy formation model \texttt{GAMESH} with cluster population synthesis codes to trace the cosmic evolution of globular clusters simultaneously with mergers of massive black holes. \\
   Our reference model, which includes prescriptions for both cluster formation and disruption depending on properties of specific galaxies, accurately reproduces the observed age-mass distribution of the Milky Way's globular clusters. We find that approximately 30\% of the globular clusters observed in our galaxy's halo may have originated from satellite galaxies of the Milky Way. We confirm that hierarchical black hole mergers provide a significant contribution to the formation of black holes in and above the pair-instability mass gap. However, quantifying their contribution is challenging, as different population synthesis codes yield divergent results in terms of the black hole mass function and merger rates. Furthermore, we characterize the host galaxies where massive black holes form in terms of their dark matter, stellar mass, and metallicity. \\
   Ultimately, we demonstrate that the merger and birth rate densities of binary black holes increase with redshift until \textit{z} = 5. This cosmic evolution is a crucial signature with significant implications for future detectors such as LISA, the Einstein Telescope, and Cosmic Explorer, which will be capable of probing the high-redshift Universe.
   }


    \maketitle

\section{Introduction}

The detection of gravitational waves (GWs) in 2015 \citep{Abbott+2016} marked the beginning of a new era in astrophysics, unveiling the mysterious universe of black holes (BHs). Since then, the LIGO-VIRGO-KAGRA (LVK) Collaboration has published  nearly 180 binary BHs (BBHs) \citep{Abbott+2023, GWTC4v1+2025}, providing crucial insights into the fundamental processes that govern the formation and evolution of these enigmatic compact objects \citep{Abbott+2017, Abbott+2019, Abbott+2021, Abbott+2023BHPop, Abbott+2024, GWTC4+2025}.\hfil\break
Despite remarkable progress being made, the properties of the merging BBHs detected by LVK, such as their mass and spin distributions, continue to challenge theoretical models. For example, the absence of a high-mass cutoff in the primary BH mass distribution above $\sim 60\, \rm M_\odot$ \citep{Farmer+2019, Farmer+2020, VanSon+2020, Marchant+2020, Callister+2024} has raised questions about the definite existence of a pair-instability mass gap. This gap, which is predicted by stellar evolution models, does not appear in the LVK catalog \citep{Abbott+2023BHPop, GWTC4+2025}, which includes several detections of BHs within the gap region \citep{Abbott+2019}. Nevertheless, the precise astrophysical channel responsible for the formation of massive stellar BBHs (MBBHs) with $\rm M_{\rm BH} > 50\,\rm M_\odot$ remains elusive, with a variety of mechanisms proposed to explain the most massive GW events. These mechanisms include isolated binary evolution \citep{Bethe+1998, Belczynski+2008, Dominik+2012, Dominik+2013, Dominik+2015, Belczynski+2016, Schneider+2017, Giacobbo+2018, Marassi+2019, Graziani+2020b, Broekgaarden+2022, Franciolini+2024}, direct star collisions \citep{DiCarlo+2019, Renzo+2020, Kremer+2020a, Kremer+2020b, Costa+2022, Ballone+2023}, three-body interactions \citep{Antonini+2017, ArcaSedda+2021, VignaGomez+2021}, and dynamical formation channels involving star clusters \citep{Mapelli+2016, Kumamoto+2020} and the disks of active galactic nuclei \citep{McKernan+2012, Bartos+2017, McKernan+2018, Tagawa+2019, Tagawa+2020, Ishibashi+2020, Vaccaro+2024, McKernan+2024}.\hfil\break
Among the proposed mechanisms, dynamical formation in dense stellar environments - such as globular clusters (GCs) and nuclear star clusters - has attracted significant interest over the past few decades. These environments are particularly well suited for the formation of hierarchical BH merger chains, where the compact remnants of massive stars can undergo successive coalescences, leading to the formation of increasingly massive BHs \citep{Miller+2002, Holley-Bockelmann+2008, Giersz+2015, Antonini+2015, Antonini+2016, Mapelli+2021, Mapelli+2022, Chattopadhyay+2023,Kritos:2022non, Torniamenti+2024}.
Indeed, due to mass segregation, BHs gradually migrate toward the cluster core on sub-gigayear timescales, resulting in the formation of a BH-dominated central region \citep{Spitzer+1969, Kulkarni+1993, Sigurdsson+1993}. In such BH-dominated cores, dynamically hard BBHs are promptly formed through three-body interactions, and further hardened by GW capture, binary-single, and binary-binary scattering events \citep{Phinney+1991, Portegies+2000, Peuten+2016, ArcaSedda+2018, Zocchi+2019, Kremer+2020a, Antonini+2020}.
Additionally, dynamical mechanisms can produce unique features that could help to distinguish the contribution of this formation channel from that of isolated binary evolution, such as large spin magnitudes inherited from the premerger orbital angular momentum \citep{Gerosa+2017} and non-negligible eccentricities \citep{Nishizawa+2016, Breivik+2016, Nishizawa+2017, Samsing+2019, Martinez+2020, Rom+2024}. \hfil\break However, the dynamical formation of BBHs presents several challenges.  A major obstacle is the relativistic kick imparted to merger remnants at birth \citep{Campanelli+2007, Lousto+2011}, which can exceed the escape velocity of their host clusters. This often results in the ejection of the remnants from their environment, thereby disrupting the hierarchical merger chain \citep{Favata+2004, Holley-Bockelmann+2008, Kesden+2010}. The interplay between the properties of the host environment and the strength of the kicks plays a critical role in shaping the mass distribution of the BH population resulting from this formation pathway~\citep{Islam:2025drw}. \hfil\break
Unfortunately, due to the large distances involved and limited sky localization accuracy, confidently determining the orbital properties and specific birth environments of BBHs remains challenging. This uncertainty hampers our ability to fully understand the true origin of MBBHs and to rule out alternative formation channels, such as the binary evolution of Population~III stars \citep{Kinugawa+2014, Kinugawa+2016, Tanikawa+2021b,  Wang+2022, Tanikawa+2022, Costa+2023, Santoliquido+2023, Mestichelli+2024, Liu+2024} and primordial BHs \citep{DeLuca+2021, Ng:2022agi, ChenZu+2024, DeLuca+2025}
as contributors to the observed GW events.\hfil\break
As the number of detected mergers continues to grow with ongoing and future runs, the prospect of disentangling the contributions from different formation mechanisms becomes increasingly attainable. Bayesian inference methods \citep{Thrane+2019, Abbott+2023BHPop, Franciolini+2022, Franciolini+2024} and population synthesis models \citep{Hurley+2000, Portegies+1995, Mapelli+2013, Spera+2015, GiacobboMapelli+2018, Elbert+2018, Breivik+2020, Riley+2022, Iorio+2023} enable the inference of the intrinsic properties of observed BBHs by comparing observational data with predictions derived from various astrophysical scenarios and binary initial conditions. These statistical comparisons have opened new avenues for uncovering the true formation history behind these mergers, despite the significant uncertainties associated with stellar and binary evolution \citep{Iorio+2023}. \hfil\break
In this work, we focus on the role of the dynamical formation scenario in dense stellar environments, with particular emphasis on the contribution of BBHs formed in GCs to the observed population of BH mergers. Additionally, we investigate how the distinctive properties of BHs resulting from hierarchical mergers may provide new insights into the astrophysical processes underlying the formation of MBBHs.\hfil\break
To investigate this, various approaches have been employed. Traditionally, the numerical complexity of direct $N-$body and Monte Carlo simulations has limited studies of cluster dynamics to a relatively small number of cases \citep{Spurzem+1999, Rodriguez+2015, Rodriguez+2016a, Rodriguez+2016b, Wang+2015, Wang+2016, Chatterjee+2017, Askar+2017, Rodriguez+2019, Banerjee+2020, ArcaSedda+2023, ArcaSedda+2024}. However, the recent development of fast semi-analytical codes, also known as cluster population synthesis (CPS) codes -- such as 
\texttt{B-POP} \citep{ArcaSedda+2021}, \texttt{cBHBd} \citep{Antonini+2020,Antonini+2023},
\texttt{FASTCLUSTER} \citep{Mapelli+2021}, 
\texttt{QLUSTER} \citep{Gerosa+2023}, and
\texttt{RAPSTER} \citep{Kritos+2024} -- has made it possible to conduct cluster population studies within a reasonable time-frame. By combining CPS predictions with models of the cosmic cluster formation rate \citep{Mapelli+2022, GonzalezPrieto+2022, Kritos+2024, Torniamenti+2024, Mestichelli+2024}, one can rapidly estimate the average merger rate density of dynamically formed BBHs as a function of redshift. However, this approach does not provide detailed information about the spatial distribution or the chemical properties of the galaxies hosting globular clusters.\hfil\break
Understanding the properties of host galaxies is crucial for improving our knowledge of GCs and, by extension, of MBBH formation \citep[see][for a review]{Krumholz+2019}. Several key aspects of cluster formation are known to vary with the galactic environment, including the cluster formation efficiency \citep{Ginsburg+2018}, the potential high-mass truncation of the cluster initial mass function \citep{Wainer+2022}, and the mass–radius relation \citep{Brown+2021}. Moreover, the time evolution of individual star clusters is strongly influenced by a range of environmental factors \citep[e.g.,][]{Rossi+2016, Suin+2022}. Therefore, a realistic model of GC formation and evolution must incorporate the evolving, galaxy-specific environmental conditions, and do so across a statistically significant sample of galaxies to adequately represent their diversity.\hfil\break
By incorporating the progression of chemical enrichment in galaxies throughout their assembly, hydrodynamic galaxy formation simulations offer a powerful tool to study the population of dynamically formed BBHs emerging from dense stellar environments. However, due to their high computational cost, these simulations require a trade-off between mass and spatial resolution, the size of the simulated volume, and the redshift range covered. Currently, most high-resolution zoom-in simulations -- those with sub-parsec resolution capable of capturing star cluster formation -- are limited to small cosmological volumes and/or high redshifts \citep[e.g.,][]{Boley+2009, Kimm+2016, Lahen+2019, Calura+2022, Calura+2024, Sugimura+2024}. When the mass resolution is insufficient to resolve star cluster formation within galaxies, semi-analytical models become essential to complement these simulations. Studies utilizing this approach have already been carried out, demonstrating promising agreement with the observed GC populations in nearby galaxies \citep[e.g.,][]{Li+2014, Pfeffer+2018}. \hfil\break
This study aims to investigate the physical properties of dynamically formed BBHs originating in GCs during the assembly of the Local Group (LG) simulated by the galaxy formation model \texttt{GAMESH} \citep{Graziani+2017}. By estimating their merger rate density and examining their birth environments, we quantify the contribution of this dynamical formation channel to the GW events detected by the LVK Collaboration. Additionally, our analysis enables the identification of the host galaxies of these BBHs and the characterization of their physical properties. To this end, in Section~\ref{sec:GalaxyFormation} we outline the adopted galaxy formation model, in Section~\ref{sec:Coupling} we detail the coupling between the CPS framework and the galaxy evolution simulation, and in Section~\ref{sec:CPS} we describe our cluster evolution models. Our main findings are presented in Section~\ref{sec:results}, followed by an extended comparison with previous studies in Section~\ref{sec:comparison}. This discussion is especially timely because of the recent release of O4 data~\citep{LIGOScientific:2025slb}.

\section{Methods}
\label{sec:Methods}
In this study, we adopt a LG-like volume simulation, named \texttt{GAMESH} \citep{Graziani+2015} (Section~\ref{sec:GalaxyFormation}) and propose a novel semi-analytic coupling approach for GC formation (Section~\ref{sec:Coupling}), enabling the placement of cluster populations within each galaxy. Finally, we integrate the time evolution of GCs using the \texttt{RAPSTER} and \texttt{FASTCLUSTER} codes (Section~\ref{sec:CPS}), marking their first application within the context of a numerical galaxy formation simulation.

\subsection{Galaxy formation simulation: \texttt{GAMESH}} \label{sec:GalaxyFormation}
The galaxy formation model \texttt{GAMESH} \citep{Graziani+2015, Graziani+2017, Graziani+2020} combines an N-body simulation -- which traces the assembly of dark matter -- with a semi-analytic model that predicts star formation and the evolution of baryonic matter through processes such as chemical enrichment and radiative transfer.\footnote{It is worth noting that, to enable a direct comparison with the results of \citet{Graziani+2020}, we adopt the same galaxy formation run, employing a semi-analytic prescription for radiative feedback, rather than executing the numerical radiative transfer component of the pipeline.}\hfil\break
Our semi-numerical approach focuses on a cube-shaped region measuring 4 comoving Mpc (cMpc) per side, centered on a halo analogous to the Milky Way (MW). This volume, designed to resemble the LG, is sufficiently large to encompass a statistically significant population of dwarf galaxies, enabling us to test our semi-analytic prescriptions for GC formation across a diverse range of astrophysical environments. The simulation achieves a mass resolution of approximately $3.4\,\times \, 10^5\,\rm M_\odot$, allowing us to track the dynamics of mini-halos composed of few tens of dark matter particles. \hfil\break
The N-body simulation tracks the formation of cosmic structures from redshift $z = 20$ down to the present day ($z = 0$), with snapshots saved at intervals of $15\,\rm Myr$ for $z \geq 10$, and every $100\,\rm Myr$ for $z<10$. In each snapshot, dark matter halos are identified using a Friends-of-Friends algorithm with a linking length of 0.2 and a minimum of 100 particles per halo. A particle-based merger tree is then constructed to trace the hierarchical assembly of these halos over cosmic time.\hfil\break
\texttt{GAMESH} incorporates several key baryonic processes. The mass of gas within each halo is initially assumed to be proportional to its dark matter mass, scaled by the cosmic baryon fraction. As halos evolve along the merger tree, they acquire gas through both smooth accretion from the surrounding environment and mergers with other structures. Consequently, the gas accretion rate is determined by the growth of the dark matter halo: as a halo increases its mass, it self-consistently acquires a proportional amount of cosmic gas \citep{Graziani+2015}. \\
In each halo, the star formation rate (SFR) is computed as SFR = $\epsilon_\star$ $\frac{\rm M_{gas}}{\rm t_{dyn}}$, depending on the star formation efficiency $\epsilon_\star$, the total gas mass $\rm M_{gas}$ and the dynamical time of the host halo $\rm t_{dyn} \sim (1+z)^{-\frac{3}{2}}$ \citep{Graziani+2017}. We adopt a star formation efficiency of $\epsilon_\star=0.02$, a value that has been calibrated within the \texttt{GAMESH} framework to match the observed properties of the LG. As shown in Fig. 1 of \citet{Graziani+2017}, this efficiency allows the model to successfully reproduce the cosmic SFR density evolution, ensuring that the predicted growth of stellar mass is consistent with observations at \textit{z} = 0.\\
Mechanical feedback is regulated by supernova explosions, which drive galactic winds capable of ejecting gas into the intergalactic medium. The mass of the ejected gas is calculated as $\Delta \rm M = 2\,\epsilon_{W}\,\frac{E_{SN}}{v_{esc}^2}$, where $E_{SN}$ is the energy released by supernovae, $\rm v_{esc}$ is the escape velocity of the halo, and $\epsilon_{W}$ represents the wind efficiency, which controls how effectively supernova energy is converted into kinetic energy to drive outflows \citep{Graziani+2017}.\\ 
We model the chemical enrichment of the interstellar medium by implementing a simplified supernova feedback prescription, following \citet{Salvadori+2008MNRAS}. In line with the instantaneous recycling approximation \citep{Tinsley+1980}, we assume that metals and gas are mixed both immediately and uniformly \citep{Salvadori+2007, Salvadori+2010MNRAS}.\\
Radiative feedback is treated following the simplified approach in \citet{Graziani+2017}, by assuming an instantaneous reionization redshift at \textit{z} = 6.\hfil\break
Because our simulation focuses on a volume representative of the LG, the rates of star and cluster formation within it tend to be higher than the cosmic average. Consequently, when comparing our results to studies based on larger, more representative cosmological volumes \citep[e.g.,][]{Hopkins+2014, McAlpine+2016, Kaviraj+2017, Donnari+2019, Venditti+2023}, our predictions for certain quantities should be interpreted as upper-limit estimates \citep{Mapelli+2018, Artale+2019, Broekgaarden+2022, Franciolini+2024, Bruel+2024, Liu+2024}. Finally, a key advantage of testing our novel coupling approach within a MW-like galaxy is the exceptional availability of observational data. The existence of comprehensive GC catalogs for the MW enables detailed and rigorous comparisons with our simulation outcomes \citep{Forbes+2010, Dotter+2010, Dotter+2011, VandenBerg+2013}, providing a robust framework for validating the model’s ability to reproduce the fundamental properties of galactic GCs.

\subsection{Coupling setup} 
\label{sec:Coupling}
Our coupling approach involves several key steps. First, we identify galaxies that are potential sites for GC formation. We then assign initial properties to the newly formed GCs and use these as input conditions for the CPS codes, which integrate the dynamical evolution of each cluster over time. Finally, we evaluate whether GCs are disrupted by galaxy mergers or dissolved within galactic disks. These steps are described in more detail below:

\begin{enumerate}
    \item 
    Zoom-in simulations \citep{Ma+2020, Chen+2021, Fukushima+2021} suggest that GC formation is most likely to occur when the gas surface density of giant molecular clouds exceeds approximately $10^3\,\rm M_\odot\,\rm pc^{-2}$ \citep[see][for reviews]{Adamo+2020, Kruijssen+2025}. Recent JWST observations of lensed galaxies have provided empirical support for this scenario, revealing the presence of ultra-compact proto-globular clusters with stellar surface densities reaching $\sim10^5\,\rm M_\odot\,\rm pc^{-2}$ \citep{Adamo+2024Natur}. Within the \texttt{GAMESH} framework, we estimate the gas surface density by dividing the total gas mass $\rm M_{\rm g}$ contained in a dark matter halo by the effective area of the galaxy
    \begin{equation}
      \Sigma_{\rm g} = \frac{\rm M_{\rm g}}{\pi\,\rm r^2_{\rm h}} > 10^3 \,\rm M_\odot\,{\rm pc}^{-2},  
    \end{equation}
    where $\rm r_{\rm h}$ is the half-mass stellar radius. The exact gas surface density threshold for GC formation remains a subject of debate. Hosting galaxies are expected to be highly fragmented \citep{Fujimoto+2025NatAs}, implying that local density peaks can trigger GC formation even when the disk-averaged density remains below the threshold. Although we apply this gas surface density threshold on galactic scales, the approximation is particularly valid at high redshifts, where $\rm r_{\rm h}$ is comparable to the typical size of giant molecular clouds \citep[$\sim 10-200$ pc,][]{Dame+2001}. At lower redshifts, where these scales diverge more significantly, the impact on our results remains limited, as the bulk of GC formation occurs at high redshifts (\textit{z} > 3). It is important to stress that recent high-resolution JWST observations of lensed high-redshift galaxies have discovered several bound clusters within regions smaller than 100 pc \citep{Bradac+2025ApJ, Messa+2025A&A}, confirming that at cosmic dawn the GC progenitors form at scales comparable to the typical size of giant molecular clouds.\\
    To compute $r_{\rm h}$ from the stellar mass $M_{\star}$ of each dark matter halo, we use the mass–size relation from \citet{Shen+2003}, incorporating its redshift evolution as described in \citet{vanDerWel+2014}. This scaling is consistent with recent JWST observations confirming that galaxies at \textit{z} $>$ 6 are significantly more compact than their local counterparts \citep{Mowla+2024Natur, Kalita+2025MNRAS}.\\We have performed a sensitivity test on the choice of this threshold and the results are discussed in Appendix \ref{app:GasSurfaceDensityAnalysis}.

    \item For each galaxy that can host GC formation, we compute the cluster formation efficiency $\Gamma$ \citep{Li+2017, Pfeffer+2019, Lahen+2020, Li+2022, Grudic+2023} following the work of \citet{Kruijssen+2012}:
    \begin{equation}
        \Gamma = \left(1.15 + 0.6\,\Sigma_{\rm SFR}^{-0.4}+0.05\,\Sigma^{-1}_{\rm SFR}\right)^{-1},
    \end{equation}
    where $\Sigma_{\rm SFR} = \frac{\rm SFR}{\pi\,\rm r_h^2}$ is the star formation rate surface density. A fraction of the stellar mass that forms between two snapshots, $\Delta \rm M_\star = {\rm SFR} \, \Delta t$, is assumed to be bound in GCs ($\rm M_{\rm cl}$):
    \begin{equation}
        \rm M_{\rm cl} = \Gamma\, \Delta \rm M_\star\,.
    \end{equation}

    \item We sample GC masses from a cluster initial mass function modeled as a power law with an exponent of \(-2\) \citep{Li+2017, Li+2018, Pfeffer+2019, Li+2022, Lahen+2024}. This choice is further supported by recent JWST observations of star clusters across cosmic time (\textit{z} $\approx$ 1-9), which find a mass function slope of $-1.89^{+0.13}_{-0.12}$, consistent with a universal power-law index of \(-2\) \citep{Claeyssens+2026arXiv260116281A}. \\
    Since the CPS prescriptions are typically less accurate for lighter clusters, the minimum GC mass is commonly set at \(m_{\rm cl, min} = 10^{5}\,\rm M_{\odot}\). However, for the scope of this study, we extend this minimum mass to \(10^{4}\, \rm M_{\odot}\), based on the assumption that clusters with masses below \(10^{5}\, \rm M_{\odot}\) do not contribute significantly to the overall BBH merger rate. The maximum mass, conversely, depends on the properties of the host galaxy, approximately scaling as \citep{Li+2022}:
    \[
    m_{\rm cl, max} = \Sigma_{\rm SFR}^{2/3}.
    \]
    We randomly sample GC masses from the cluster initial mass function until the remaining stellar mass bound in GC falls below \(m_{\rm cl, min}\). The initial metallicity of GCs is assumed to be the same of their parent galaxy. Their initial half-mass radii are drawn from a uniform distribution ranging between 0.1\,pc and 1\,pc \citep{Askar+2017, Kremer+2020a, ArcaSedda+2020}.
    Although these initial conditions are informed by both simulation results and observational data, a significant uncertainty remains regarding the binary fraction within GCs \citep{Richard+2015, Stanway+2020}. In this study, we adopt a binary fraction of unity for all clusters and leave a detailed exploration of its effects for future work.
 
   \item Once formed within galactic disks, GCs are not immune to disruption. They are frequently subjected to strong tidal forces from dense gas clouds, as demonstrated in various studies \citep{Gieles+2006, Elmegreen+2010, Kruijssen+2011}. Following the framework presented by \citet{Kruijssen+2015}, we model this rapid disruption process -- referred to as the ''cruel cradle effect'' -- as:

    \begin{equation}
        \left(\frac{dm_{\rm cl}}{dt}\right)_{\mathrm{cce}} = -\frac{m_{\rm cl}}{t_{\mathrm{cce}}},
    \end{equation}\label{eq:cce}

    where $t_{\mathrm{cce}}$ denotes the characteristic timescale of the cruel cradle effect \citep{Kruijssen+2012b}. This timescale depends on various galactic properties and is expressed as \citep{Kruijssen+2015}:

    \begin{equation}
    t_{\mathrm{cce}} = 176\,\left(\frac{f_\Sigma}{4}\right)^{-1}
    \left(\frac{\rho_{\mathrm{ISM}}}{\rm M_\odot\,\mathrm{pc}^{-3}}\right)^{-3/2}
    \left(\frac{m_{\rm cl}}{10^5\,\rm M_\odot}\right)\,
    \phi_{\mathrm{ad}}^{-1}\, \mathrm{Myr},
    \end{equation}

    where:

    \begin{itemize}
    \item $f_\Sigma = 3.92\,\left(\frac{10 - 8\,f_{\mathrm{mol}}}{2}\right)^{1/2}$, with $f_{\mathrm{mol}} = \left(1 + 0.025\,\Sigma_{\rm g}^{-2}\right)^{-1}$;
    
    \item $\rho_{\mathrm{ISM}} = \frac{3\,\Omega^2(r_{\rm g})}{\pi\,G}$ is the average mid-plane density of the interstellar medium in a stable galactic disk \citep{Krumholz+2005, DeLucia+2024}, evaluated at the galactocentric radius $r_{\rm g}$ of the GC. In this study, we assume GCs to be on circular orbits and located at a radius $r_{\rm g} = r_{\rm h}$. Although this represents a simplified orbital model, we perform a dedicated sensitivity analysis in Appendix \ref{app:OrbitalAssumptions} to assess the impact of this assumption on our results and to quantify any potential systematic effects;
    
    \item $\phi_{\mathrm{ad}} = \left[1 + 9\,\left(\frac{\rho_{\rm h}/\rho_{\mathrm{ISM}}}{10^4}\right)\right]^{-3/2}$ is a correction factor accounting for the tidal energy absorbed by adiabatic expansion. Here, $\rho_{\rm h} = \frac{3\,m_{\rm cl}}{8\pi\,r_{\mathrm{cl}}^3}$ is the cluster's internal density, and $r_{\mathrm{cl}}$ is the cluster radius.
    \end{itemize}

    \item Finally, we apply a GC survival prescription during galaxy mergers, following the models presented in \citet{Kruijssen+2011, Kruijssen+2015, DeLucia+2024}. After the coalescence of two galaxies, we compute the fraction $f_{\mathrm{surv}}$ of GCs that migrate from the galactic disks of their progenitor galaxies - where they typically form - to the halo of the merger remnant, as described in \citet{Kruijssen+2012b}:

    \begin{equation}
        f_{\mathrm{surv}} = 4.5 \times 10^{-8} \left(\frac{m_{\mathrm{cl,min}}}{100\,\rm M_\odot}\right)^2 
        \left(\frac{t_{\mathrm{depl}}}{\mathrm{yr}}\right)^{0.77 - 0.22 \log(m_{\mathrm{cl,min}}/100\,\rm M_\odot)},
    \end{equation}\label{eq:galmerg}

    where $t_{\mathrm{depl}}$ is the gas depletion timescale, estimated as the ratio of the gas mass to the peak star formation rate (SFR) during the merger-induced starburst \citep{Kruijssen+2012b}. Following \citet{DeLucia+2024}, we approximate this peak SFR by summing the SFRs of the two merging galaxies at the snapshot in which the merger is identified.

    Once relocated to the halo, GCs are assumed to evolve in isolation and are unaffected by any subsequent mergers \citep{Lamers+2010}. The migrating GCs are randomly selected from those residing in the progenitor galaxies' disks prior to the merger.

    The survival fraction depends critically on the type of merger. In minor mergers (baryonic mass ratio $< 0.1$), only the GCs from the less massive galaxy migrate to the halo, while those in the more massive galaxy remain in its disk. In major mergers, all GCs from both progenitors are assumed to migrate.\\
    We clarify that the disruption mechanisms described by Equations \ref{eq:cce} and \ref{eq:galmerg} are applied to the initial cluster population, before their evolution is tracked via the CPS models. Specifically, this preliminary analysis allows us to determine the lifetimes of GCs. Once this preliminary selection is made, the subsequent mass evolution tracked within the CPS models focuses exclusively on the mass loss induced by the external galactic potential.
\end{enumerate}

\subsection{Cluster population synthesis codes: \texttt{RAPSTER} and \texttt{FASTCLUSTER}}
\label{sec:CPS}

To generate large populations of GCs and study their dynamically formed BBHs, the only practical approach both in terms of computational efficiency and statistical robustness is to employ CPS codes such as \texttt{RAPSTER} \citep{Kritos+2023, Kritos+2024} and \texttt{FASTCLUSTER} \citep{Mapelli+2021, Torniamenti+2024}.
These tools adopt a semi-analytic framework specifically designed to model the coupled evolution of massive star clusters ($m_{\mathrm{cl}} \geq 10^5\,\rm M_\odot$) and their resident BBH populations.
The main advantage lies in the ability to simulate a large number of clusters efficiently, a critical feature for statistical studies of dynamically formed BBHs within cosmological simulations.\\
The codes are built upon a set of physically motivated prescriptions calibrated against direct N-body and Monte Carlo simulations. These describe all relevant dynamical channels, such as three-body interactions, GW captures, and binary-single scatterings. Their simplified prescriptions circumvent the computational intensity of direct $N$-body simulations, allowing to evolve individual GCs with $m_{\mathrm{cl}} \geq 10^5\,\rm M_\odot$ within a matter of minutes. However, the computational speed comes at a cost: only the average properties of the cluster and its BHs are evolved with time, whereas the internal dynamics of the cluster is not integrated directly. By employing both \texttt{RAPSTER} and \texttt{FASTCLUSTER}, we aim to quantify the impact of different semi-analytical prescriptions on our results, ensuring that our findings regarding dynamically formed BBHs remain robust across different modeling approaches within the same cosmological framework.\\
Their considerable flexibility enables to explore a wide range of input parameters, including BBH masses, spins, initial orbital eccentricities, delay times to merger, and various properties of the host star clusters, such as their masses, densities, and initial binary fractions. Furthermore, these CPS codes allow for the temporal evolution of the host stellar cluster itself, incorporating key astrophysical processes such as stellar mass loss, two-body relaxation, and tidal stripping by the host galaxy. \hfil\break
While both codes are effective CPS tools, their main differences lie in their primary focus. 
With its fast Monte-Carlo sampling algorithm and extensive calibration on GC data, \texttt{FASTCLUSTER} aims to reproduce statistically time-dependent observations, such as the merger rate of BBHs when coupled with analytic star formation histories. Conversely, \texttt{RAPSTER} is a more unconstrained model with several free parameters, initially developed to investigate the fundamental possibility of producing massive and intermediate-mass BHs from globular and nuclear star clusters. Despite these differences, both codes are specifically optimized for GW astronomy, with the shared goal of creating extensive catalogs of dynamically formed BBH mergers for comparison with observational data. For further information on how the implementation of this physics takes place, we refer to the main papers for each code.\hfil\break
To compute the initial single-BH mass spectrum, both codes adopt the stellar evolution prescriptions from \texttt{SEVN} \citep{Spera+2017, Iorio+2023}, which determine the remnant mass as a function of the zero-age main-sequence mass of stars. This study focuses exclusively on GCs composed of Population~II and I stars; thus, we sample initial stellar masses from a Kroupa initial mass function (IMF) \citep{Kroupa+2001}, spanning the range $[0.08,\,200]\,\rm M_\odot$ and covering metallicities in the interval $[10^{-2}, 1]\,Z_\odot$. Given that our LG environment is overdense, metal enrichment in low-metallicity halos occurs extremely rapidly. As a consequence, only a few halos reach the conditions necessary for Population~III GC formation. Their contribution to the overall results is negligible and, hence, not considered here; we defer their treatment to future work focused on high-redshift regimes. \\
Although we adopt the same initial conditions for GCs, intrinsic differences between the working principle of CPS codes persist. Preliminary tests performed with non-calibrated code settings revealed significant structural differences in the algorithms, leading to substantial divergences in final outcomes. Specifically, variations in setting parameters such as the initial binary fraction and the number of BH merger chains within each GC were identified as major sources of discrepancy.\\
To address these code-specific structural differences and avoid highly divergent results, we implement a calibration method focusing on the most critical diverging element, i.e., the BH merger chains.  More precisely, \texttt{FASTCLUSTER} requires the number of BH merger chains within each GC as an explicit input parameter, a value that \texttt{RAPSTER} does not explicitly demand but intrinsically calculates. We therefore extract this value from the \texttt{RAPSTER} outcomes and adopt the identical number in \texttt{FASTCLUSTER} for an accurate one-to-one comparison.

\section{Results}
\label{sec:results}

Here we present the main results of our study. In Section~\ref{sec:GC formation} we examine the formation and destruction of GCs and their properties in terms of age and metallicity. In Section~\ref{sec:BHB results} we explore the mass distribution of dynamically formed BBHs originating within GCs, with particular attention to their birth environments (Sec. \ref{sec:FormationEnvironments}). In Section~\ref{sec:Merger Rate} we report the BBH birth and merger rate densities as a function of redshift. Finally, in Section~\ref{sec:comparison} we compare our predictions with previous studies.

\subsection{Properties of globular clusters and comparison with observations}
\label{sec:GC formation}
In this section we present the physical properties of GCs in our simulations, including their distributions in mass, metallicity, and age. We compare these characteristics to current observational data for MW GCs \citep{Kruijssen+2015}.

\begin{figure*}[h!]
    \centering
    \begin{subfigure}[b]{0.50\textwidth}
        \includegraphics[width=\textwidth]{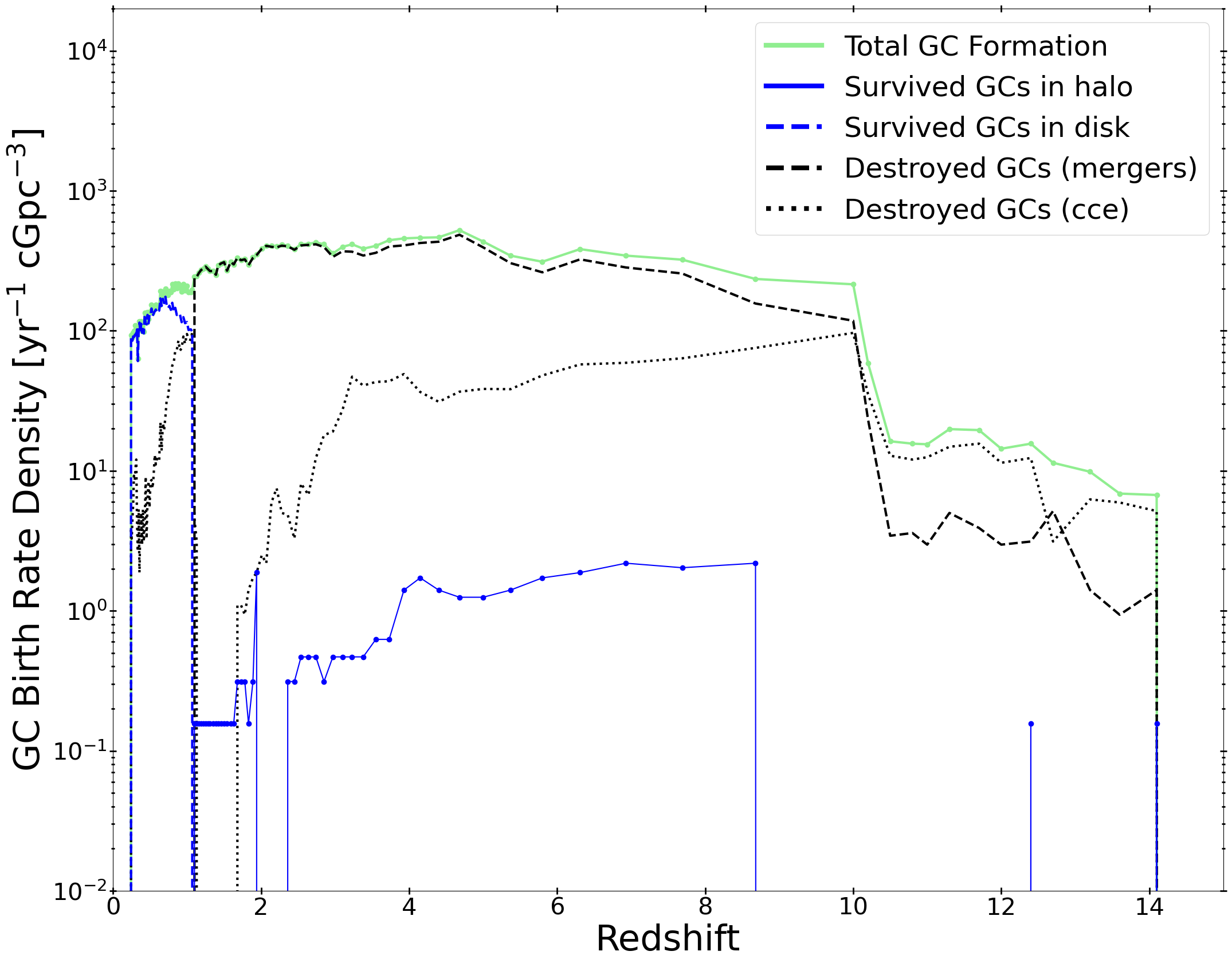}
        \caption{}
        \label{fig:GCFormation}
    \end{subfigure}
    \begin{subfigure}[b]{0.48\textwidth}
        \includegraphics[width=\textwidth]{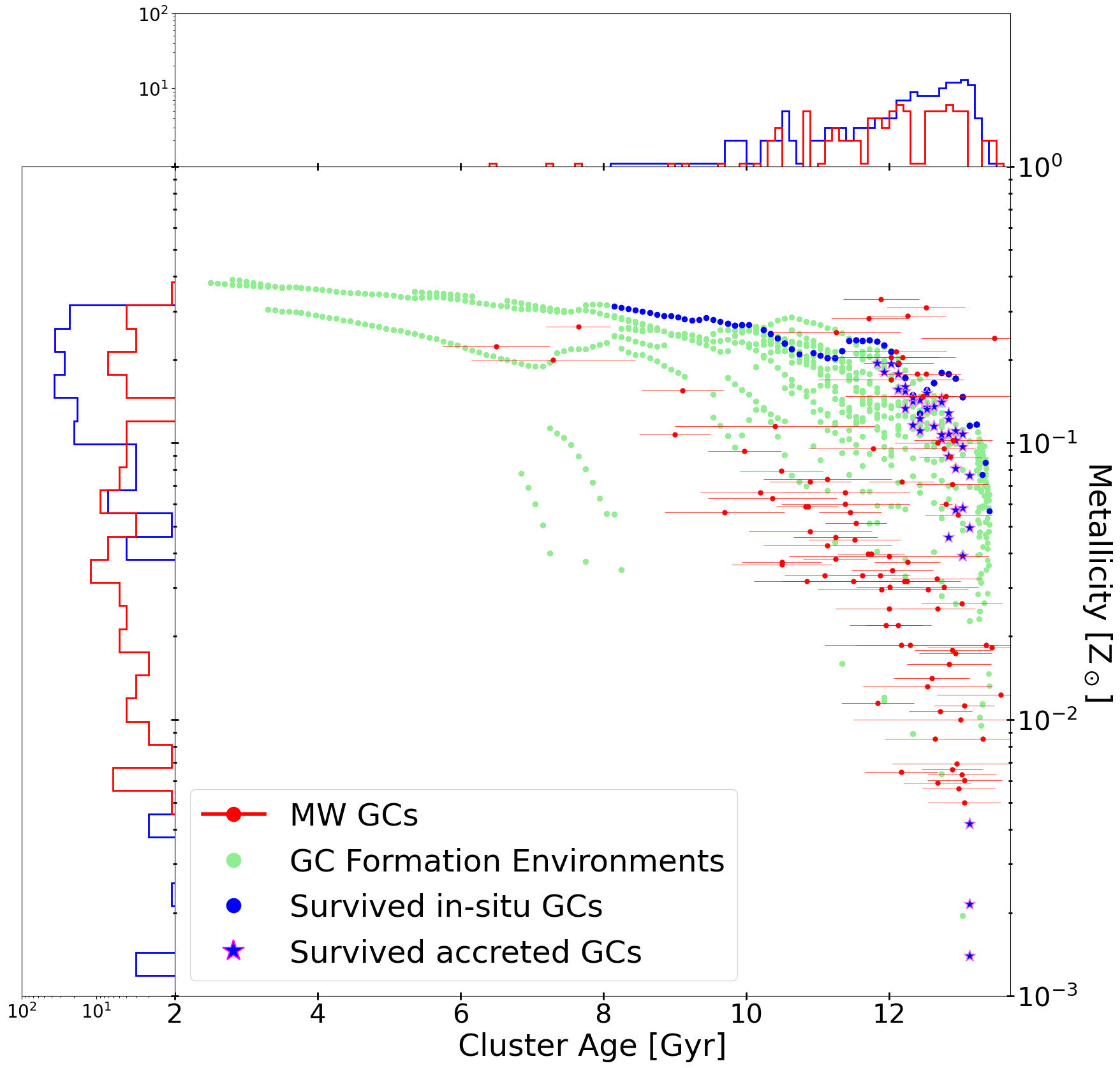}
        \caption{}
        \label{fig:AgeMetallicityGC}
    \end{subfigure}
    \caption{(a) Global birth rate density of GCs in the \texttt{GAMESH} simulation as a function of redshift (solid green line). Black lines represent GCs that will not reach \textit{z}=0 due to either galaxy mergers (dashed line) or gas interactions (dotted line), whereas blue lines show GCs that survive either in the MW halo (solid) or in the disk (dashed) until today. (b) Age-metallicity distribution of surviving MW GCs formed in situ in the MW galaxy main progenitors (blue dots) and accreted through galaxy mergers (blue stars). Red dots with error bars correspond to individual GCs observed in our Galaxy \citep{Kruijssen+2015}. Conversely, green dots represent all GC formation environments within our LG-like volume.}
\end{figure*}

We begin by examining the GC birth rate density within our LG-like volume, as shown by the solid green line in Fig~\ref{fig:GCFormation}. Our simulation distinguishes among three GC populations based on their fate: those destroyed by galactic merger events, those destroyed by interactions with giant molecular clouds, and those that survive to $z = 0$. As expected, excluding these destruction mechanisms would significantly overestimate the number of surviving GCs. The surviving population is further categorized by location, as some GCs remain in the MW halo and others reside in its disk.\\
Interestingly, our simulation reveals a peak in the total GC formation at slightly higher redshift (\textit{z} $\sim 4.5$) compared to previous studies \citep[\textit{z} $\sim 3$;][]{Bruel+2024, DeLucia+2024} that focus on cosmologic boxes.\footnote{\citet{Bruel+2024} use the FIRE simulation, which evolves a cube volume of (22.1 cMpc)$^3$, while \citet{DeLucia+2024} consider a
fraction (about 10\%) of the entire volume of the Millennium
Simulation \citep{Springel+2005}.} This earlier peak likely stems from the overdense nature of our simulated region, which promotes an earlier onset of both star and GC formation than in average-density environments. \\
The surviving GC population in the MW halo exhibits a multi-peaked distribution. The peaks are a direct result of major mergers with Lyman-$\alpha$ cooling halos, which efficiently bring a large number of GCs into the MW halo. Conversely, the smooth trend that connects these peaks is primarily driven by continuous, minor mergers with mini-halos. Despite the apparent misalignment with the single formation peak inferred from observed GCs in the MW halo (at $z\sim 3$), our results are in broad agreement with observations when considering the age uncertainties. The typical error bars on observed GC ages, ranging from 0.25 to 0.50 Gyr, are large enough to make our simulated peaks consistent with the single observed peak.\hfil\break
The dominant mechanisms for GC destruction also show a clear redshift dependence. At high redshift (\textit{z} > 10), the cruel cradle effect is the primary driver of disruption. The compact nature and smaller physical sizes of early galaxies result in frequent and highly disruptive interactions with giant molecular clouds. In contrast, at 1 < \textit{z} < 10, the influence of galaxy mergers surpasses that of gas interactions, becoming the leading cause of GC disruption. As redshift decreases toward \textit{z} $\sim 1$,  a notable fraction of GCs begin to survive within the disk of the MW main progenitor. This trend reflects the decline in the merger rate and in efficiency of the cruel cradle effect, which is driven by a significant decrease in the interstellar medium (ISM) density within the galaxy's half-mass radius at later times. The sharp decrease in the fraction of disrupted GCs at z $\approx$ 1 is primarily driven by the transition of the MW progenitors -- the main contributors to GC formation in our LG-like volume -- into a secular evolution phase, where the absence of major mergers  and the post-reionization suppression of star formation in satellites significantly reduces the efficiency of GC destruction.\hfil\break
Figure \ref{fig:AgeMetallicityGC} presents the age-metallicity distribution of GCs surviving until redshift \textit{z} = 0 in the halo of our MW-like galaxy. They exhibit a significant agreement with the estimated lifetimes of observed GCs. Nevertheless, three of the youngest observed objects deviate from our predictions. On the basis of the age determination, these GCs are predicted to form at \textit{z} < 1 when major mergers - which are responsible for introducing GCs into the MW halo - no longer occur in our simulation. It is crucial to acknowledge, however, that the substantial uncertainties in current GC age estimates preclude a conclusive determination of their true formation times. Indeed, taking into account the error bars, two of these GCs overlap with the region of our surviving GCs. \hfil\break
Our model struggles to reproduce the oldest metal rich GCs. This is because the oldest GCs are generally expected to have low metallicities, reflecting the pristine conditions of the early Universe. The metallicities of these objects, instead, are comparable to formation environments typically found at lower redshifts (cluster age $\sim8-9$ Gyrs), which is challenging to explain within our current framework. \hfil\break
Our simulated GCs tend to have systematically higher values of the metallicity ($\sim 10^{-1}$ Z$_\odot$), approximately one order of magnitude higher than the bulk of observed GCs ($\sim 3\,\times 10^{-2}$ Z$_\odot$). This discrepancy is evident when comparing the simulated data (represented in blue) with the observational data in Fig.~\ref{fig:AgeMetallicityGC}. A direct quantitative comparison of metallicity is not straightforward. Observational studies typically report metallicity as [Fe/H], whereas the \texttt{GAMESH} simulation defines metallicity as the abundance of elements heavier than helium \citep{Graziani+2015}. This fundamental difference in the definition of metallicity likely contributes to the offset between our simulation results and observations. Beyond this difference, the offset could also be influenced by how metals are diluted within the ISM in our simulation. Specifically, our model assumes that supernova-driven outflows eject gas and metals in proportions identical to the ISM, thus not altering ISM metallicity. However, these outflows might actually be more metal rich than the surrounding ISM. Furthermore, our simulation assumes homogeneous accretion of gas at the average intergalactic medium metallicity. On the other hand, enrichment processes might be far more inhomogeneous, leading to a preferential accretion of unenriched gas, which would dilute the ISM metallicity.\hfil\break
Understanding the origin of the MW GCs is another important aspect. We simultaneously present on the same Fig.~\ref{fig:AgeMetallicityGC} all GC birth environments (green dots), systems that satisfy the gas surface density threshold and have a nonzero cluster formation efficiency. This reveals a wide range of environments, including those that are metal poor (Z $\sim10^{-2}$ Z$_\odot$). We observe that the majority of survived GCs (blue dots) follow a well-defined metallicity pattern consistent with their birth in the MW main progenitor. Conversely, those deviating from this path originate in satellite galaxies that later merged with the MW. Their distinct metallicity distributions provide a clear way to distinguish accreted GCs from those native to our Galaxy. Furthermore, there is an interesting population of newly formed GCs with metallicity below 0.01 Z$_\odot$. This population is born within dwarf galaxies located at the periphery of our LG-like volume. These smaller systems, being more isolated and having lower star formation efficiencies, remain more chemically pristine compared to the main hosts. This environmental diversity reflects the bimodal metallicity distribution observed in the plot. The gap between these two populations is also sensitive to the adopted minimum gas surface density threshold for GC formation of $10^3$ M$_\odot$ pc$^{-2}$: removing this threshold would indeed reveal
lower-metallicity environments, as shown in Appendix \ref{app:GasSurfaceDensityAnalysis}.\\
Additional properties of the survived GCs, including their mass-metallicity distribution, are presented in Appendix \ref{app:MassDistribution}. This analysis reveals that our GC sample is representative of the entire spectrum of observed masses.

\subsection{Population of detectable dynamically formed binary black holes}
\label{sec:BHB results}
\begin{figure*}[h!]
    \centering
    \includegraphics[scale=0.28]{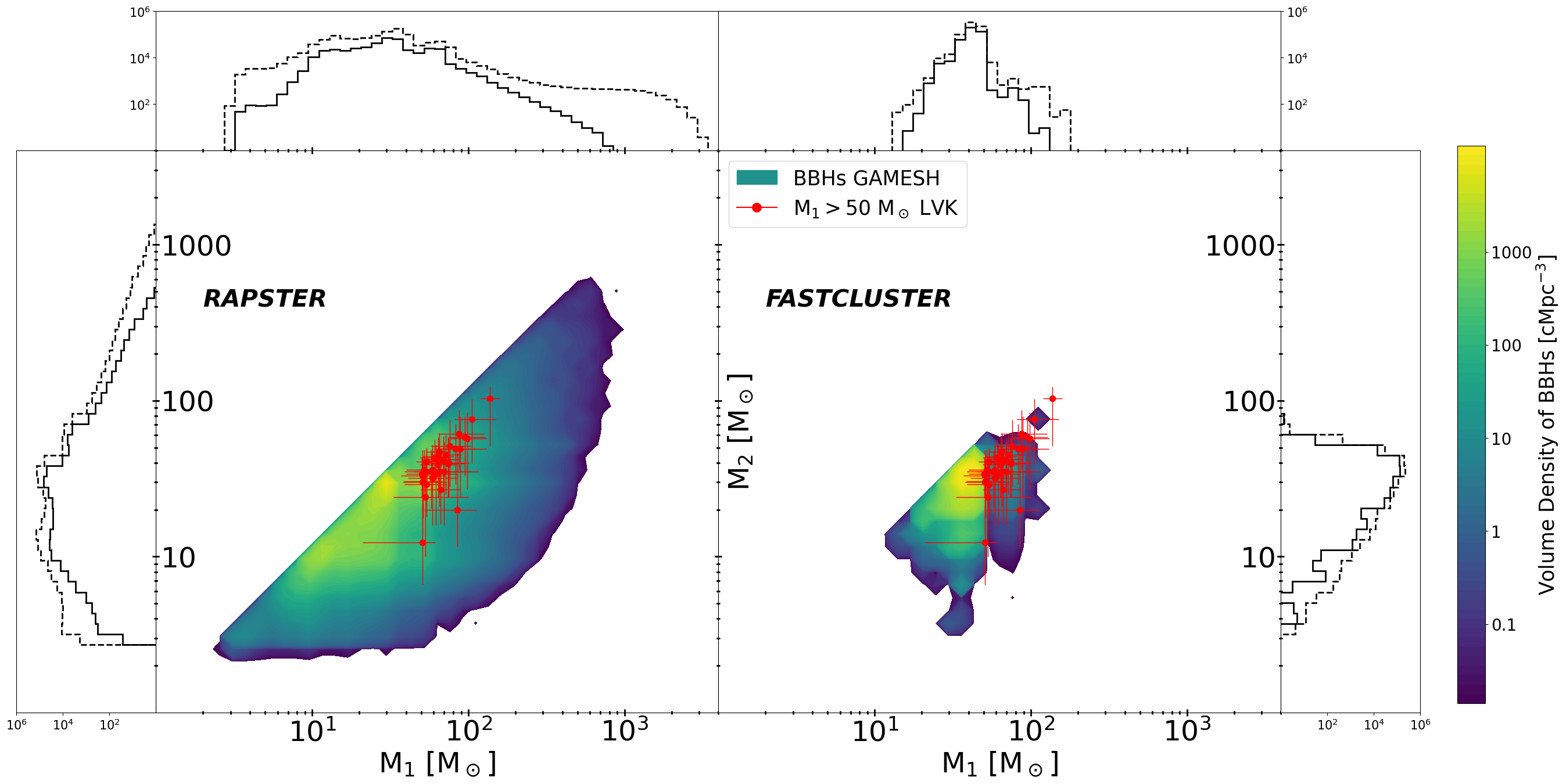}
    \caption{Primary (M$_1$) and secondary (M$_2$) BH mass distribution of observable dynamically formed BBHs within our LG-like volume. 
    Red dots show the estimated median  values of the most massive GW events (M$_{1}> 50$ M$_\odot$) detected by the LVK collaboration up to O4a, with error bars representing the 90\% credible intervals. Dashed lines in histograms display the mass distributions of all BBH mergers that occur in our simulation.
    }
    \label{fig:M1M2Plot}
\end{figure*}

In this section we provide an in-depth analysis of dynamically formed BBH mergers during our LG's assembly, focusing on those detectable by current terrestrial interferometers. \hfil\break
Figure~\ref{fig:M1M2Plot} shows the mass distribution of primary (M$_1$) and secondary (M$_2$) BHs of dynamically formed BBHs in GCs. 
The color code in the density map represents the volume density of BBH mergers in our LG. Brighter (darker) regions indicate higher (lower) event occurrence densities. To assess their detectability, we adopt GWFish \citep{Dupletsa+2023}, a dedicated Python package based on Fisher matrices for GW population studies. We set a signal-to-noise ratio threshold of 8 to identify events that would be observable by the current network of LVK detectors \citep{GWTC4v2+2025}.
\hfil\break
The primary BH mass distribution offers several insights into the origin of the LVK detections, because different CPS codes yield distinct predictions for the mass range of dynamically formed BHs. The \texttt{FASTCLUSTER} code struggles to explain GW events with M$_1 <15$ M$_\odot$ through dynamical formation processes. In contrast, such events are plausible, albeit with lower probability densities, in \texttt{RAPSTER}.\\
An interesting aspect of the M$_1$ distribution is the presence of two prominent peaks in the distribution predicted by both CPS. 
The lower-mass peak, predicted by both CPS codes around M$_1 \sim 35$ M$_\odot$, appears also in BPS codes but is significantly less prominent in those models \citep{GiacobboMapelli+2018, Belczynski+2020, Tanikawa+2020, Broekgaarden+2022, Iorio+2023}. The key difference, indicative of the hierarchical formation scenario, is the presence of the second peak (M$_1\sim 70-90$ M$_\odot$) that is a distinguishing feature of the CPS predictions. The presence of MBBHs (e.g., M$_1>50$ M$_\odot$) within the predictions of both codes strongly suggests a dynamical origin for these massive systems. \hfil\break
A tail of high-mass mergers extends even into the regime of IMBHs (M$_{\rm BH}\sim10^2-10^4$ M$_\odot$) in the \texttt{RAPSTER} simulations, but not in \texttt{FASTCLUSTER}.
Despite the formation of thousands of IMBHs in our \texttt{RAPSTER} simulations, just tens of them are potentially detectable by current GW observatories with very low signal-to-noise ratio. This is due to the high masses and merger redshift of these systems, which will be shown in detail in Section~\ref{sec:Merger Rate}. \\
\begin{figure*}[h!]
  \centering
    \begin{subfigure}[b]{0.49\textwidth}
        \includegraphics[width=\textwidth]{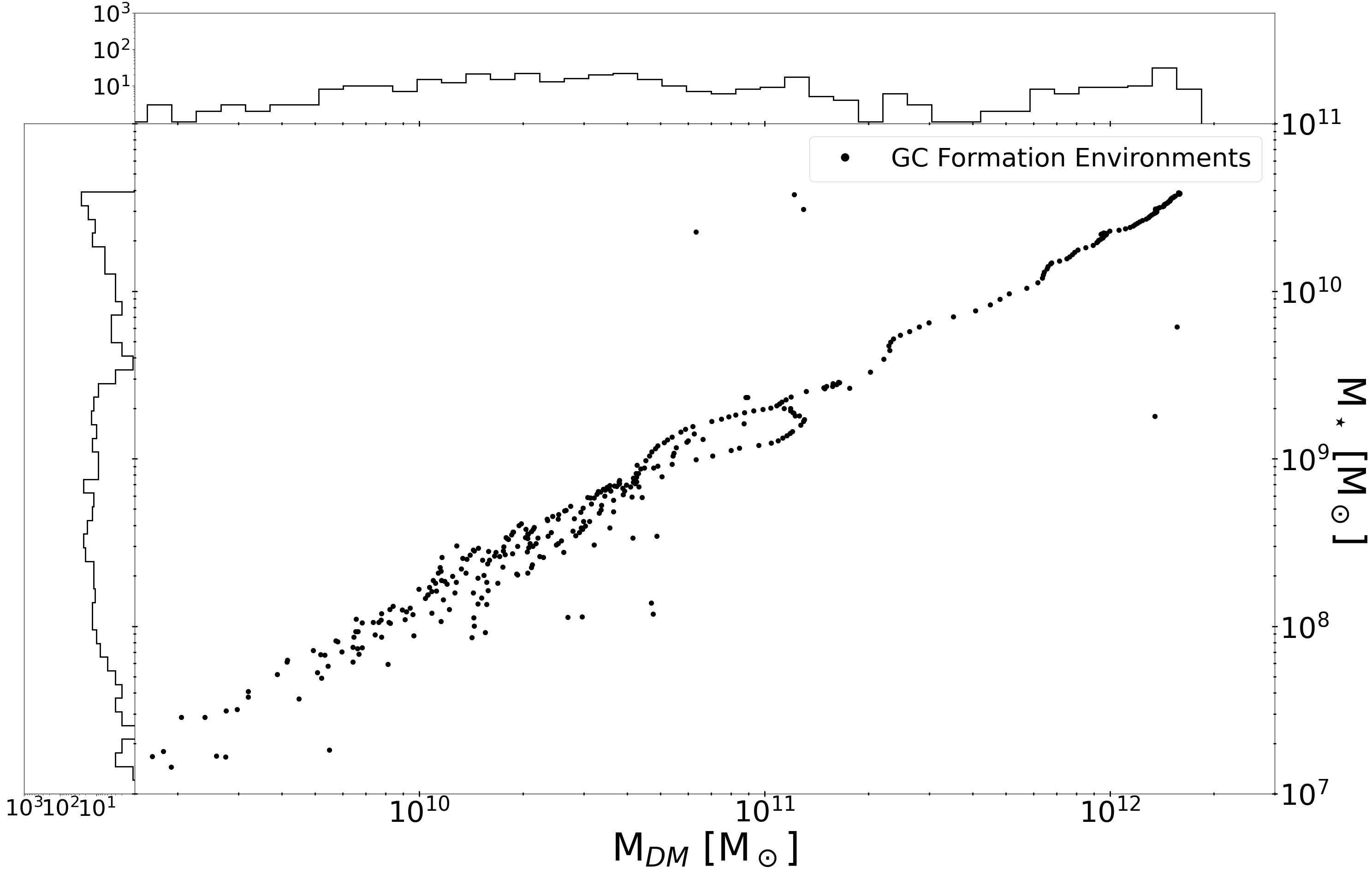}
        \caption{}
        \label{fig:HostingGalaxies}
    \end{subfigure}
    \begin{subfigure}[b]{0.50\textwidth}
        \includegraphics[width=\textwidth]{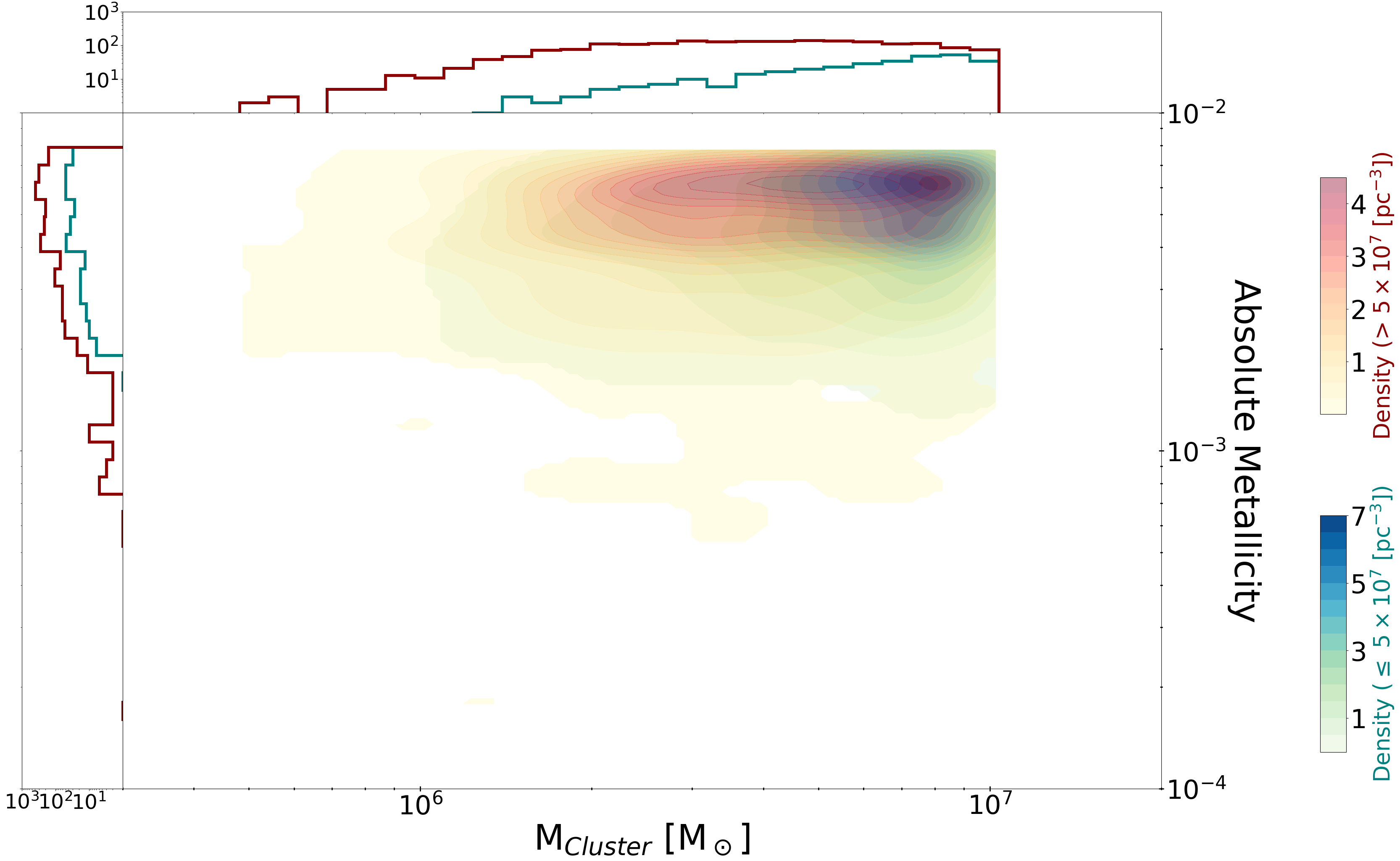}
        \caption{}
        \label{fig:HostingGCs}
    \end{subfigure}
    \caption{(a) Dark matter mass (M$_{DM}$) and stellar mass (M$_\star$) distribution of halos that host GCs capable of forming MBBHs. (b) Two-dimensional density distribution of GCs in the mass-metallicity plane. The contours represent the density estimation of the cluster population, highlighting the regions where initial central stellar number density and total mass facilitate MBBH formation, with color gradients distinguishing between high initial central number density (red-yellow tones, $\rho_c >$ $5\times10^7$ pc$^{-3}$) and lower initial central density (blue-green tones, $\rho_c <$ $5\times10^7$ pc$^{-3}$). }
    \label{fig:Hosts}
\end{figure*}
The location of IMBHs is an important aspect that our approach is able to provide. Due to their high masses, GW kicks are not large enough to eject these IMBHs from their host stellar clusters. Most of these clusters do not survive as they are destroyed by interactions with gas or by galaxy mergers. Consequently, the IMBHs that once resided in such clusters become ''wandering IMBHs'' within galaxies. Galaxy mergers are chaotic dynamical processes and it is challenging to precisely trace the final position of the BHs. however, in the case of GCs destroyed by the cruel cradle effect, it is plausible that their IMBHs remain confined to the disk. Here, due to dynamical friction \citep{Capuzzo2008, ArcaSedda+2014}, they will tend to migrate toward the galactic center, potentially contributing to the formation of the super-massive BH in the nucleus. The last column of Table \ref{Tab:Statistics_BHBs} shows the number of IMBHs that remain confined to the MW disk. These BHs can grow through mergers to masses spanning from $10^3$ M$_\odot$ to $4\times 10^3$ M$_\odot$. This mass range represents a lower limit, as our work does not account for any gas accretion processes. \\
Only a minority of IMBHs form within GCs that survive until \textit{z} = 0. These IMBHs, remaining bound to their host clusters, could therefore be observable in the halo of our Galaxy. The last column of Table \ref{Tab:Statistics_BHBs} shows, in parenthesis, the number of IMBHs that migrate into the Galaxy's halo together with their GCs.  In the halo, they can grow to a mass range of $\sim2\times 10^3$ M$_\odot$. So, this work provides crucial insights into the expected distribution and potential observational locations of IMBHs, guiding future observational surveys. A more detailed discussion of this topic will be presented in future papers.\hfil\break
Table \ref{Tab:Statistics_BHBs} presents the statistics of merging BBHs from the \texttt{GAMESH} simulation. As the table shows, the number of BBH mergers is comparable between the two CPS codes and this consistency persists even for systems observable by terrestrial interferometers. By comparing the first and second column, we see that a significant population of BBHs remains undetectable. This happens for several key reasons, primarily related to the frequency of the GWs produced and the sensitivity of the detectors. 
A major limiting factor is the specific frequency range of terrestrial detectors, such as LIGO, Virgo, and KAGRA, which are mostly sensitive from approximately 10 Hz to a few kHz. Many BBH mergers, especially those involving very massive BHs (M$_{\rm BH} > 100$ M$_\odot$) or those in their early inspiral phase, produce GWs at frequencies well below this sensitivity range.\\
Future observatories, such as LISA, Cosmic Explorer and the Einstein Telescope, are being developed to overcome the sensitivity limitations of the current network. These detectors will be sensitive to a much lower frequency range (mHz) and higher redshift, allowing them to detect more massive BHs and to provide a more complete picture of the BBH population.\\
The third column of Table \ref{Tab:Statistics_BHBs} indicates the number of merging and detectable BBHs that originate in the MW. It is clear that most of the merging systems form within our Galaxy, with only a small fraction (10\%) originating in MW satellites. 

\subsection{Properties of galaxies and globular clusters hosting massive binary black hole formation}
\label{sec:FormationEnvironments}

Figure~\ref{fig:HostingGalaxies} illustrates the properties of the halos that host GCs capable of forming MBBHs. The log-log plot displays a strong correlation between the dark matter mass (M$_{DM}$) and the stellar mass (M$_\star$) of these environments. This relationship, which spans several orders of magnitude, closely follows the well-established stellar mass-halo mass relation observed for galaxies \citep{Moster+2013}. The tightness of this correlation suggests that the formation of these specific GCs is not a stochastic process, but it is fundamentally linked to the overall mass and evolutionary state of their host galaxy.\\
The histograms along the axes provide further insight into the mass distributions. The top histogram shows that the majority of these environments have a dark matter mass between $10^{10}-10^{12}$ M$_\odot$, a range that includes also progenitors of the MW. The corresponding stellar mass distribution is shown in the left histogram, and it does not exhibit a significant peak that would suggest a preferred range of values.\\
\begin{figure*}[t]
    \centering
    \begin{subfigure}[a]{0.80\textwidth}
        \includegraphics[width=\textwidth]{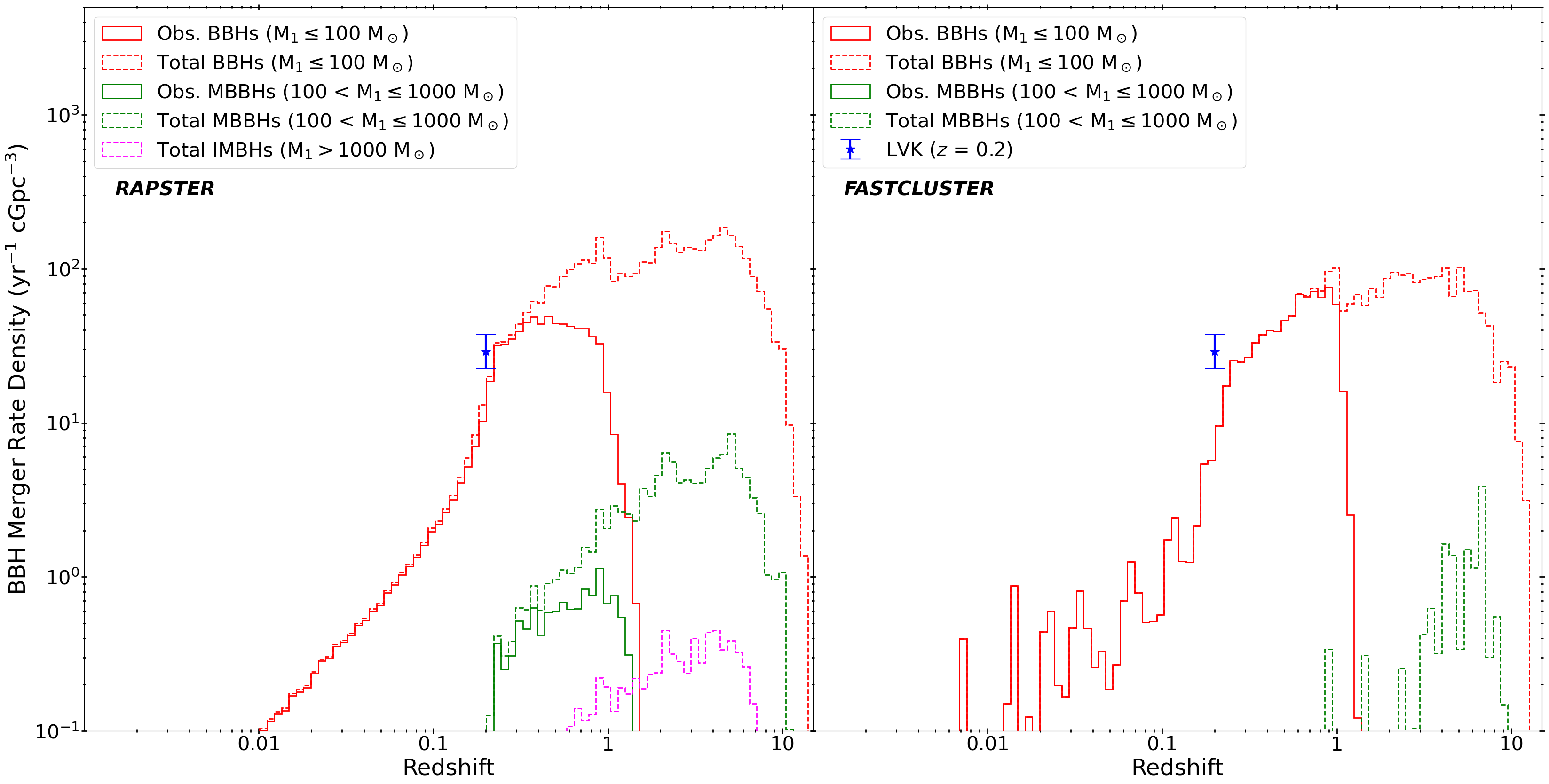}
        \caption{}
        \label{fig:BHB_MergerRateDensity}
    \end{subfigure}
    \begin{subfigure}[b]{0.80\textwidth}
    \includegraphics[width=\textwidth]{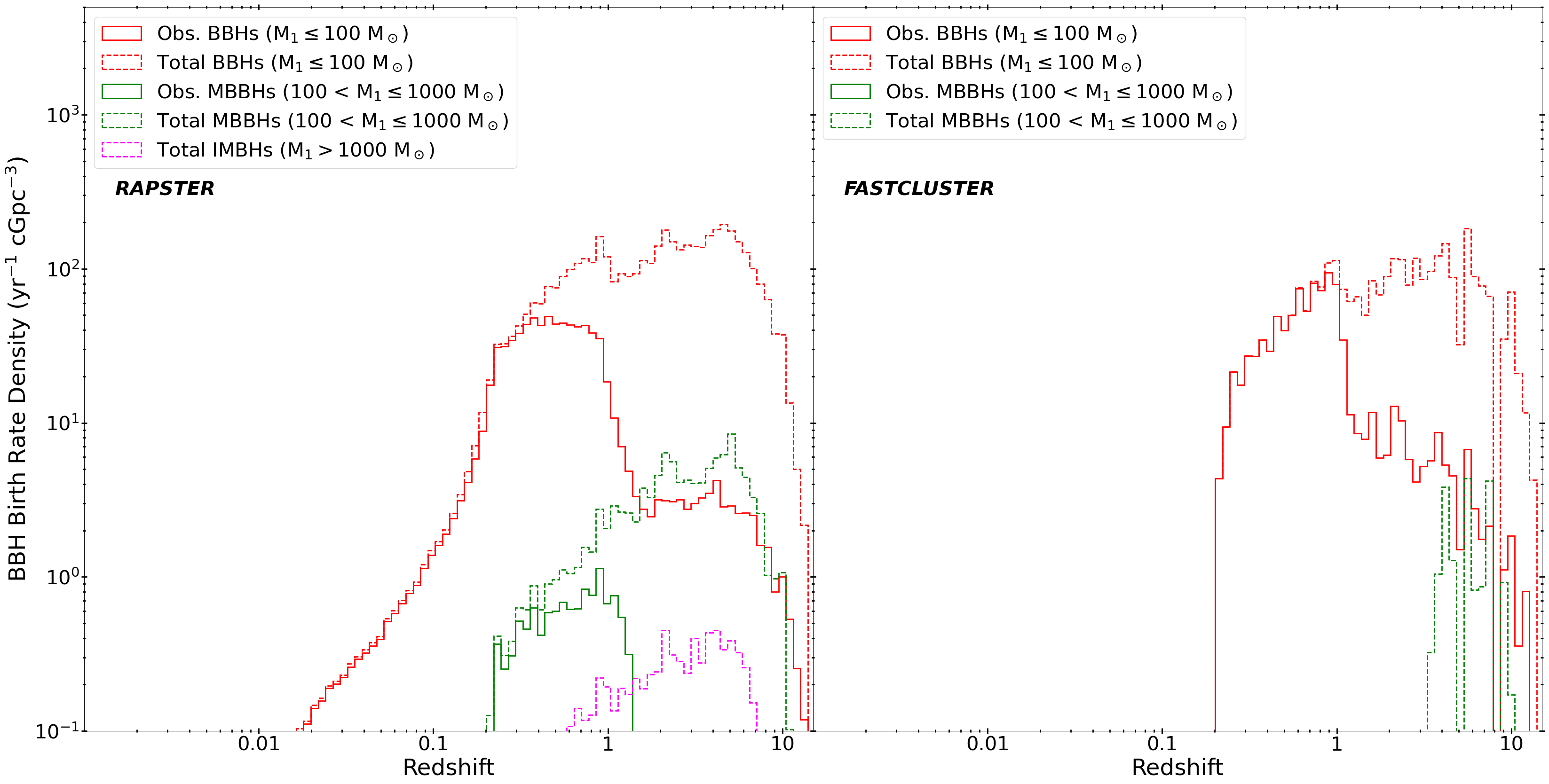}
    \caption{}
    \label{fig:BHB_BirthRateDensity}
    \end{subfigure}
    \caption{(a) Redshift evolution of dynamically formed BBH merger rate density, as a function of primary BH mass (M$_1$). Red curves show BBH with M$_1\leq100$ M$_\odot$, green lines indicate MBBHs with $100$ M$_\odot <$ M$_1\leq1000$ M$_\odot$, and violet histograms illustrate IMBHs with M$_1>1000$ M$_\odot$. Blue star with error bars represents the value provided by the LVK collaboration \citep{GWTC4v2+2025}. (b) Redshift evolution of dynamically formed BBH birth rate density, as a function of primary BH mass (M$_1$).  Red curves show BBH with M$_1\leq100$ M$_\odot$, green lines indicate MBBHs with $100$ M$_\odot <$ M$_1\leq1000$ M$_\odot$, and violet histograms illustrate IMBHs with M$_1>1000$ M$_\odot$.}
\end{figure*}
A critical feature of Fig.~\ref{fig:HostingGalaxies} is the clear lower bound, with no halos below a dark matter mass of approximately $10^9$ M$_\odot$ being able to originate GCs that form MBBHs. This implies that a minimum halo mass is a necessary condition for the physical processes required for this particular channel of GC and MBBH formation. Possible explanations include the need for a sufficiently deep gravitational potential well or a minimum gas reservoir to support the necessary star and cluster formation. In summary, while GCs form in a wide range of galactic environments, those forming MBBHs require a fundamental mass threshold, in terms of dark matter and stellar mass.\\
In Fig.~\ref{fig:HostingGCs} we show the absolute metallicity distribution of these GCs as a function of their mass, both on a logarithmic scale. The 2D contours represent the number density distribution of clusters, while the marginal histograms on the top and left axes show the 1D distributions for mass and metallicity, respectively. The color gradient indicates the initial central stellar number density ($\rho_c$): red-yellow tones represent higher density clusters ($\rho_c >$ $5\times10^7$ pc$^{-3}$), while blue-green tones correspond to lower density ones ($\rho_c \leq$ $5\times10^7$ pc$^{-3}$). \\
Most GCs have metallicities in the range of $10^{-3}-10^{-2}$ and a mass typically in the upper tail of our initial distribution, i.e., between $10^6-10^7$ M$_\odot$. Higher density clusters (red-yellow tones) tend to populate the less massive regions of the plot, while lower density clusters (blue-green tones) are found  closer to the upper mass limit. This suggests that the initial central stellar number density and the total mass of a GC play complementary roles in the formation of massive BHs. A lower central density can be compensated for by a higher mass, whereas less massive GCs require a higher core density to form MBBHs. However, the plot indicates a necessary lower limit for initial central density (> $10^6$ pc$^{-3}$) to efficiently form massive BHs through hierarchical mergers. \\
Figure~\ref{fig:HostingGCs} shows that GCs forming MBBHs are generally the most massive and densest clusters that we were able to extract starting from our initial power law distribution (Sec. \ref{sec:Coupling}). This correlation is significant because it connects the two key requirements for MBBH formation. Indeed, in a high-density environment, stars and BHs are much closer together. This dramatically increases the rate of gravitational encounters and binary interactions. Moreover, the most massive clusters are the ones with the largest potential reservoirs of stellar-mass BHs. So, the high density in their cores provides the dynamic environment needed to efficiently merge these BHs into a more massive object. Consequently, the presence of massive, high-density clusters in our sample directly supports the hypothesis that the dynamical conditions necessary for forming MBBHs via hierarchical mergers are present in our simulation. \\
By combining the information from the two panels of Fig.~\ref{fig:Hosts}, we can explain why all massive GCs have such high metallicities. This is a direct consequence of the galaxy mass-metallicity relation, where more massive galaxies are more metal rich. Since our simulations show that massive GCs are exclusively hosted by galaxies that exceed a certain mass threshold, their metallicities are naturally constrained to a specific, metal rich range. This indicates that the formation of MBBHs is not exclusive to metal poor environments, but it can also occur in more metal rich environments under the right initial conditions.\\

\subsection{Redshift evolution of binary black hole merger rates}
\label{sec:Merger Rate}

Figure~\ref{fig:BHB_MergerRateDensity} illustrates the merger rate density of detectable dynamically formed BBHs as a function of redshift. These systems are categorized by the mass of primary BHs: lighter BBHs (M$_1\leq100$ M$_\odot$) are shown in red solid lines, while systems with $100$ M$_\odot <$ M$_1\leq1000$ M$_\odot$ and M$_1>1000$ M$_\odot$ are displayed in green and violet solid lines, respectively. The left panel presents results obtained using \texttt{RAPSTER}, while the right panel shows predictions from \texttt{FASTCLUSTER}. \\
\begin{table*}[h!]
    \caption{Statistics of merging BBHs and IMBHs from the \texttt{GAMESH} simulation.}
    \label{Tab:Statistics_BHBs}
    \centering
    \begin{tabular}{lcccc}
    \hline\hline
    \noalign{\smallskip}
    Model & \# BBH Mergers & \# Obs. BBH Mergers & \# Obs. BBH Mergers MW & \# IMBHs \\
    \noalign{\smallskip}
    \hline
    \noalign{\smallskip}
    \texttt{RAPSTER}     & $1.14 \times 10^6$ [$2.39 \times 10^4$] & $4.09 \times 10^5$ [$5.44 \times 10^3$] & $3.87 \times 10^5$ [$5.13 \times 10^3$] & 238 (4) \\
    \texttt{FASTCLUSTER} & $7.06 \times 10^5$ [$1.21 \times 10^3$] & $4.21 \times 10^5$ [18]                & $3.17 \times 10^5$ [15]                & 0 \\
    \noalign{\smallskip}
    \hline
    \end{tabular}
    
    \tablefoot{
    Columns indicate the total number of merging BBHs ($\rm M_1 \leq 100\,\rm M_\odot$), the number of systems observable by terrestrial interferometers, the number of events formed within the MW, and the number of IMBHs ($\rm M_1 > 1000\,\rm M_\odot$) in the MW disk. The number of MBBHs ($100$ M$_\odot < $ M$_1 \leq 1000$ M$_\odot$) is shown in square brackets, while values for IMBHs that migrated into the MW’s halo are in round parentheses.
    }
\end{table*}

The two CPS codes predict a slightly different number of detectable merging BBHs, as already shown in Fig.~\ref{fig:M1M2Plot} and Table \ref{Tab:Statistics_BHBs}. Specifically, \texttt{RAPSTER} yields a lower merger rate estimate, characterized by a prominent peak at redshift of \textit{z} $\sim0.5$ followed by an exponential decline toward \textit{z} = 0. In contrast, the \texttt{FASTCLUSTER} merger distribution peaks at \textit{z} = 1, with a value slightly higher than the \texttt{RAPSTER} peak. Declines in both plots are attributed to the gradual decrease in the formation rate of GCs over time.\\ 
\texttt{RAPSTER} predicts that merging MBBHs (with $100$ M$_\odot <$ M$_1\leq1000$ M$_\odot$) are detectable at redshift \textit{z} < 2. In contrast, \texttt{FASTCLUSTER} predicts very few detectable MBBHs, and thus their corresponding rate density is not visible on the plot (< $10^{-1}$ yr$^{-1}$ cGpc$^{-3}$).\\
To visualize the fraction of mergers not currently detectable, the distribution of all merging BBHs within a Hubble time is plotted with dashed lines, using the same color scheme to denote mass range. As expected, the majority of mergers occurring at high redshift (\textit{z} > 2) is not detectable, and this is where the distribution of our CPS models predicts the peaks.\\
Interestingly, both CPS models show that the distribution of BBHs in the mass range of $100$ M$_\odot <$ M$_1\leq1000$ M$_\odot$ peaks at \textit{z} $\sim8$. In this mass range, \texttt{RAPSTER} consistently predicts a higher rate density for these systems.\\
An evident difference between the two CPS codes is that \texttt{RAPSTER} predicts the presence of IMBHs with masses greater than 1000 M$_\odot$ that remain outside the current detectability window. The existence of these massive systems, particularly at high redshift, is significant because they could be the seed of supermassive BHs at the centers of galaxies. However, \texttt{FASTCLUSTER} does not predict the formation of these type of compact objects at all, though we are adopting the same initial conditions for both CPS codes. So, the predictions for the formation and merger rates of these systems are highly dependent on the adopted CPS model. The discrepancy in the formation of IMBHs between the two codes stems from their fundamental treatment of cluster dynamics. \texttt{FASTCLUSTER} operates as a semi-analytical framework where cluster properties (such as mass M and half-mass radius r$_h$) evolve through average differential equations \citep[e.g., ][]{Gieles+2021Nat}. This approach lacks the resolution to capture the stochastic, runaway growth of a single compact object. Consequently, the absence of IMBHs in \texttt{FASTCLUSTER} is not a numerical error, but a direct consequence of its mean-field statistical nature, as it is not designed to capture those rare cases where a single object rapidly grows within a tightly packed cluster.\\
A key factor driving the discrepancy is the sensitivity of the results to the initial binary fraction f$_b$. \texttt{FASTCLUSTER} global evolution remains largely unaffected by variations in f$_b$, whereas the formation of IMBHs in \texttt{RAPSTER} is strictly dependent on it. This behavior highlights the different physical treatments: \texttt{RAPSTER} explicitly models binary-mediated heating. In a high-f$_b$ environment, the continuous hardening of BH binaries provides the necessary kinetic energy to support the BH subsystem against rapid contraction, maintaining a high-density "active" core where hierarchical mergers can proceed. \\
The fact that IMBH formation in \texttt{RAPSTER} vanishes at lower f$_b$ confirms that the runaway growth we report is a dynamically driven process fueled by cumulative binary interactions effects that are simplified in \texttt{FASTCLUSTER} framework, which treats binary interactions as isolated events without collective dynamical feedback. A full investigation into how the initial binary fraction affects IMBH formation will be addressed in future works to completely clarify its role.  \\
Figure~\ref{fig:BHB_BirthRateDensity} (where we use the same color and line style conventions as in Fig.~\ref{fig:BHB_MergerRateDensity}) illustrates the distribution of the birth rate density for dynamically formed BBHs. A direct comparison between the birth and merger rate density distributions provides significant insights into the dynamics of these binaries. Lighter BBHs (M$_1 < 100$ M$_\odot$) consistently represent the majority of the BH population across all redshift values. Perhaps unexpectedly, some detectable BBHs originate at very high redshifts ($2 \leq $ \textit{z} $\leq 10$). This suggests that, according to both CPS models, the probability of observing systems formed in the early Universe is not negligible.\\
The strong similarity in the birth and merger rate trends for MBBHs ($100$ M$_\odot < $ M$_1 \leq 1000$ M$_\odot$) and IMBHs (M $> 1000$ M$_\odot$) indicates an exceptionally rapid formation and coalescence timescale for these dynamically formed systems. The formation of IMBHs is evident at redshifts above \textit{z} $\sim1.5$ for \texttt{RAPSTER}, whereas no IMBH formation is predicted at any redshift from \texttt{FASTCLUSTER}.\\
Our analysis clearly shows that the number of BH mergers increases at high redshift. This is a crucial finding for future space and terrestrial-based interferometers, which will be able to probe these earlier cosmological epochs and potentially detect these events.

\section{Discussion}
\label{sec:comparison}
The majority of existing works on GC formation focus on cosmological volumes \citep{Pfeffer+2019, Ma+2020, Keller+2020, Doppel+2023, Bruel+2024}, rather than on over-dense regions similar to our LG. As expected, our predictions for GC formation and abundance exceed those from cosmological simulations \citep{DeLucia+2024, Bruel+2024}. This is also seen in our GC formation rate, which is higher than the value inferred from the GWTC-3 catalog by a factor of a few tens \citep{Fishbach+2023}. This result highlights that our predictions should be considered as an upper limit when compared to values from other cosmological simulations and observations, which are limited by factors such as mass resolution and current detector sensitivity, respectively.\\
The over-prediction of GC formation in our model may be a consequence of our semi-analytic coupling framework, which lacks the gas hydrodynamics needed to accurately capture the clumpy nature of the ISM. This limitation may lead to an overestimated GC formation efficiency. \\\\
Regarding the age-metallicity relationship, our results show a marked difference from some recent studies. \citet{Kruijssen+2019, Kruijssen+2020} and \citet{DeLucia+2024}, using the \texttt{E-MOSAICS} and Millennium simulations, respectively, find that their simulated GCs are systematically younger than the observed MW GCs, with an offset of $\sim0.75$ Gyr. In contrast, our model shows good agreement with the observed age distribution of GCs. However, we find significant discrepancies in the metallicity distribution. Our results on this matter align with those of \citet{DeLucia+2024}, who concluded that the metallicity distribution of simulated GCs is highly sensitive to the treatment of chemical enrichment. This discrepancy in metallicity may stem from our model's lack of gas hydrodynamics and the resulting inability to accurately capture metallicity gradients within individual galaxies. Although our model reproduces the global properties of the MW, the discrepancy in the GC metallicity distribution suggests that our prescriptions for how metals are diluted and recycled within the ISM need to be revised for future GC formation studies. Specifically, our model may not be adequately capturing the processes that enrich the gas from which GCs form, especially in the early Universe, where most of the surviving GCs were born.\\
Stellar mergers in young clusters \citep[e.g.,][]{DiCarlo+2019} can produce gap-BHs by creating massive stars that avoid the PI limit. However, our results suggest that in GCs the formation of objects exceeding 50 M$_\odot$ is strictly tied to the cluster's structural evolution, specifically requiring a central stellar density $> 10^7$ pc$^{-3}$ and a total mass  $> 10^6$ M$_\odot$. This highlights a higher degree of "environmental specialization" compared to young clusters.\\
While primordial BHs \citep{Carr+2016PhRvD} or models with revised nuclear rates \citep{Farmer+2019} predict a distribution that it is not constrained by the chemo-dynamical evolution of the clusters, our model predicts a clear redshift dependency. The formation of MBBHs and IMBHs in our scenario peaks at z$\sim$3, following the GC formation peak. This leads to a higher frequency of mergers involving BHs $>$ 100M$_\odot$ at high redshifts (z$>$2), offering a potential observational test to distinguish our channel from others.\\
As shown in Section~\ref{sec:Merger Rate}, our two CPS codes yield different results for the BBH merger rate density ($\mathcal{R}_{BBH}$). At the LVK Collaboration's fiducial redshift of \textit{z} = 0.2, we find merger rates of $\mathcal{R}_{BBH}\sim21.2$ and 12.9 Gpc$^{-3}$ yr$^{-1}$ for \texttt{RAPSTER} and \texttt{FASTCLUSTER}, respectively. Both of these values slightly differ from the inferred value, which is in the range $22.5-37.5$ Gpc$^{-3}$ yr$^{-1}$ \citep{GWTC4v2+2025}. In any case, BBH mergers from this formation channel could be a potentially significant contribution to the total number of observed BH mergers. The difference in the merger rates between the two codes also highlights how sensitive the local merger rate density is to the specific GC evolution models.\\

\section{Conclusions}
\label{Sec:Conclusion}

The formation of MBBHs and IMBHs remains an open problem in astrophysics. In this work we investigate hierarchical BH mergers within GCs, a primary channel for their formation. Our approach couples the \texttt{GAMESH} galaxy formation model, which simulates a LG-like volume with a MW-like galaxy at the center, with two modern CPS codes, \texttt{FASTCLUSTER} and \texttt{RAPSTER}. An approach that couples CPS codes with cosmological simulations provides a more physically comprehensive and self-consistent framework than using analytical or empirical trends. This provides crucial clues about where and when MBBHs and IMBHs form, offering a more realistic representation of their formation history in the context of galactic evolution. \\
Our results from the coupling \texttt{GAMESH}-CPS simulations indicate that hierarchical mergers within GCs provide a significant contribution to the formation of BHs with masses exceeding 50 M$_\odot$. Our main findings are as follows:
\begin{itemize}
    \item The formation of MBBHs and IMBHs in GCs requires specific initial conditions. We find that the most massive BHs (exceeding 50 M$_\odot$) form in GCs with a central stellar number density greater than 10$^7$ pc$^{-3}$ and a total mass exceeding $\sim10^6$ M$_\odot$. These two properties show a complementary trend, meaning that less massive GCs must be more compact to produce these massive objects.
    \item There is no clear correlation between metallicity and the formation of massive compact objects. This indicates that even more metal-enriched GCs can contribute to the formation of these objects.
    \item MBBHs predominantly form within well-evolved dark matter halos (M$_{DM} > 10^9$ M$_\odot$) with a stellar mass exceeding 10$^7$ M$_\odot$. This finding highlights the crucial role of galactic assembly in providing the dense stellar environments necessary for the formation of these massive compact objects.
    \item The formation of MBBHs and IMBHs predominantly occurs close to the peak of GC formation, at a redshift of $z\sim 3$. Consequently, mergers involving BHs exceeding 100 M$_\odot$ become more frequent at high redshifts (\textit{z} > 2).
    \item Predictions for the formation of MBBHs and IMBHs are highly sensitive to the specific prescriptions of the CPS code used. Key factors include the treatment of mass loss from stellar evolution, mass segregation, and tidal stripping by external potentials.
    \item Different CPS codes can lead to very different mass ranges for dynamically formed BHs. \texttt{RAPSTER} predicts a mass range spanning from 4 to $\sim$ 4000 M$_\odot$, while \texttt{FASTCLUSTER} constrains the mass range to a narrower window of $15-200$~M$_\odot$. This disparity underscores the heavy dependence of model outcomes on the internal physics of the CPS code.
    \item The two CPS codes predict slightly different merger rate densities $\mathcal{R}_{BBH}$, both of which are very close to the LVK inferred value.
\end{itemize}
Future work will tackle the problem of MBBH formation by using a more sophisticated cosmological simulation, such as \texttt{dustyGadget} \citep{Graziani+2020c}, which may help to reconcile the predictions of our cluster formation model with observations.

\begin{acknowledgements}
       F.A. acknowledges support from Sapienza and Tor Vergata University of Rome. F.A. also acknowledges support from a MSCA-GRU action funded by the European Union during the visiting period (2024 October–November) at Johns Hopkins University, Baltimore (USA). This work was supported by the Piano Nazionale di Ripresa e Resilienza (PNRR). We are grateful to Angela Adamo for her insightful suggestions on globular cluster modeling. E.B. and K.K. are supported by NSF Grants No. AST-2307146, PHY-2513337, PHY-090003, and PHY-20043, by NASA Grant No. 21-ATP21-0010, by John Templeton Foundation Grant No. 62840, by the Simons Foundation [MPS-SIP-00001698, E.B.], by the Simons Foundation International [SFI-MPS-BH-00012593-02], and by Italian Ministry of Foreign Affairs and International Cooperation Grant No. PGR01167. K.K. is supported by the Onassis Foundation - Scholarship ID: F ZT 041-1/2023-2024. MM and ST acknowledge financial support from the European Research Council for the ERC Consolidator grant DEMOBLACK, under contract no. 770017 and from the German Excellence Strategy via the Heidelberg Cluster of Excellence (EXC 2181 - 390900948) STRUCTURES. ST acknowledges financial support from the Alexander von Humboldt Foundation for the Humboldt Research Fellowship. We have benefited from the publicly available programming language \texttt{Python}, including the \texttt{numpy}, \texttt{matplotlib}, and \texttt{scipy} packages.
\end{acknowledgements}

\bibliographystyle{aa}
\bibliography{References}

\begin{appendix}
\section{Survived globular cluster mass-metallicity distribution}
\label{app:MassDistribution}

\begin{figure}[h!]
    \centering
    \includegraphics[scale=0.13]{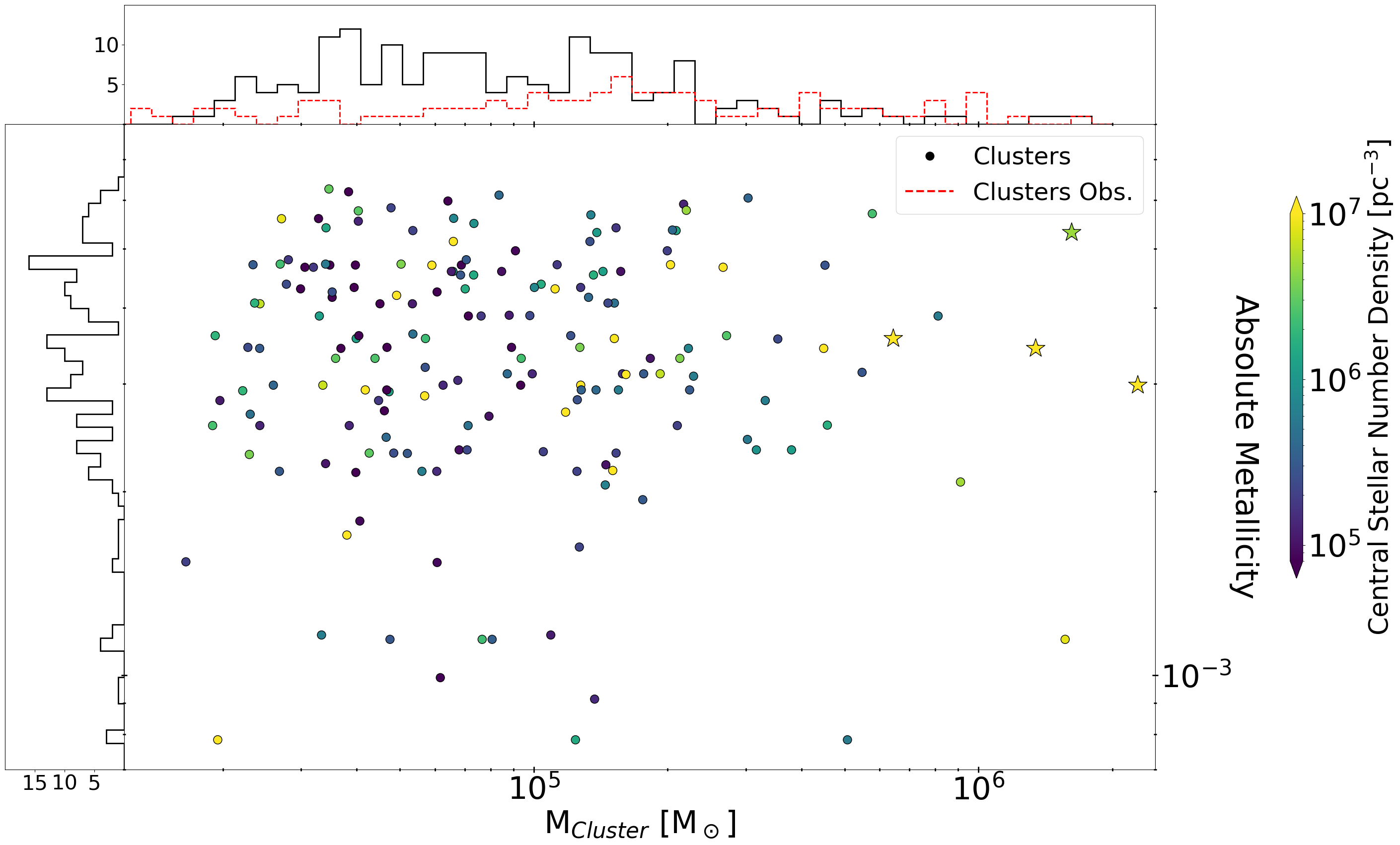}
    \caption{Mass-metallicity distribution of surviving MW GCs. The color bar indicates the initial central stellar number density of GCs, while star-shaped symbols mark the GCs hosting an IMBH. The dashed red line in the top histogram highlights the mass distribution of the observed GCs in our MW \citep{Kruijssen+2015}.
    }
    \label{fig:GCMassDistribution}
\end{figure}

Figure \ref{fig:GCMassDistribution}  illustrates the mass-metallicity distribution of GCs that survive to redshift \textit{z} = 0. The number and final masses of these GCs are in good agreement with observational data, as shown by the significant overlap between the simulated and observed GC mass distributions in the top histogram. The majority of surviving GCs experience substantial mass loss, typically around one order of magnitude, given their initial masses exceeding $10^5$ M$_\odot$. While our sample generally covers the observed mass range, we note some difficulty in populating the  extreme ends of the mass spectrum. Specifically, the under-representation of the highest masses may imply a need for more massive initial clusters in our seeding population. Conversely, for the less massive GCs, it is likely that these samples originated from initial masses below $10^5$ M$_\odot$, thereby falling outside the GC mass range considered in our study.\hfil\break
The filled points in Fig.~\ref{fig:GCMassDistribution} are color-coded by the initial central stellar number density of these GCs, spanning a wide range of values ($10^{5}-10^7$ pc$^{-3}$). This broad distribution confirms that our selection process of surviving GCs introduces no discernible bias toward a limited range of initial densities. Interestingly, despite the presence of highly compact GCs, we find that only four IMBHs form in the MW halo. This finding suggests that high compactness alone is insufficient for IMBH formation; a sufficiently high initial cluster mass (around $10^6$ M$_\odot$) is also a necessary condition (see Section~\ref{sec:FormationEnvironments}). Higher masses are crucial for refueling hierarchical BH merger chains, while higher compactness promotes BH encounters and accelerates IMBH formation. The GCs hosting the formation of IMBHs have been marked with stars in the plot. \hfil\break

\section{Sensitivity of globular cluster lifetimes to orbital assumptions}
\label{app:OrbitalAssumptions}
\begin{figure}[h!]
    \centering
    \includegraphics[width=0.47\textwidth]{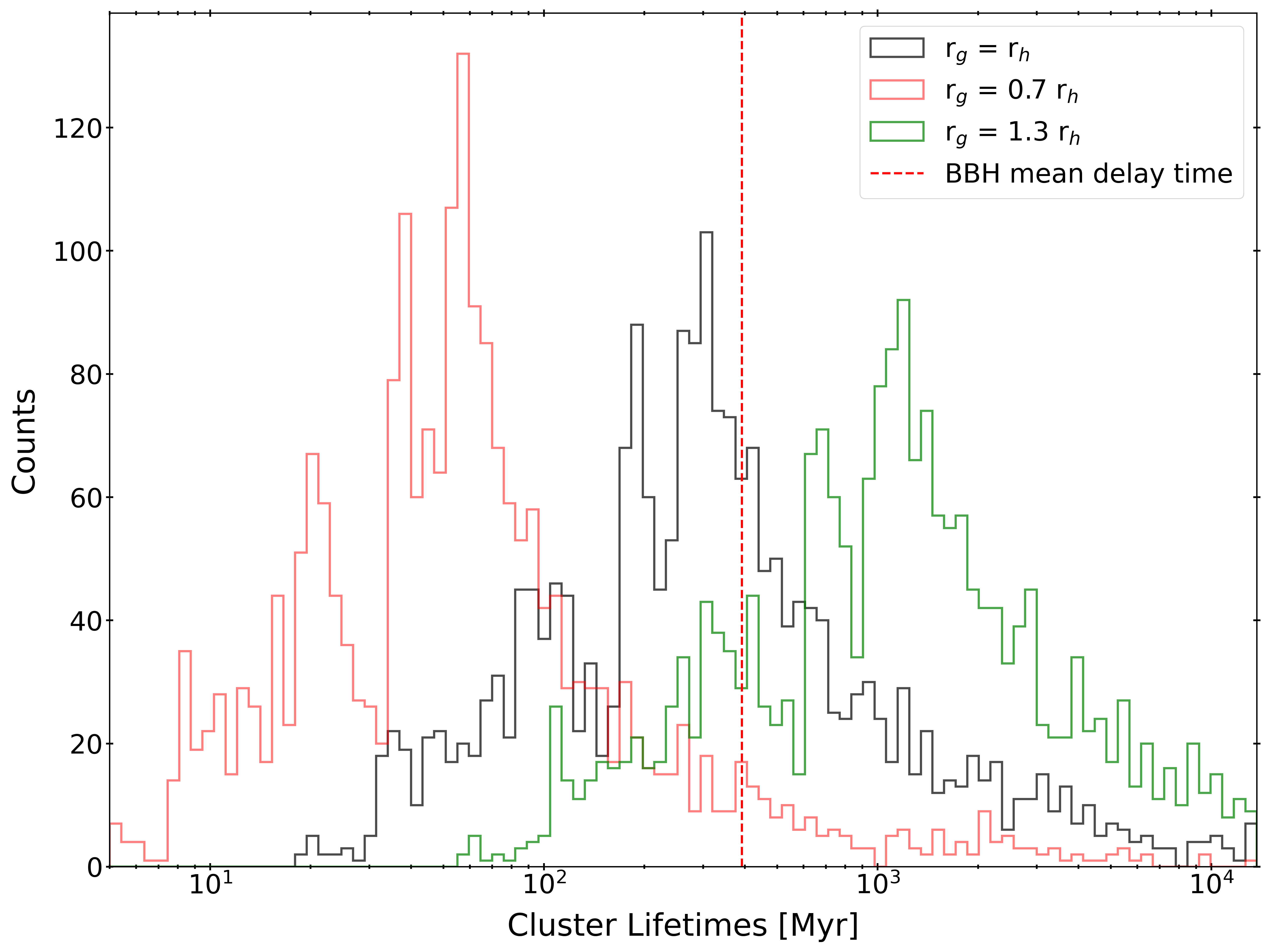}
    \caption{Distributions of GC lifetimes as a function of their orbital distance from the galactic center. Black line represents our fiducial model (r$_g$ = r$_h$), while red and green histograms indicate deeper orbits (r$_g$ = 0.7 r$_h$) and more peripheral orbits (r$_g$ = 1.3 r$_h$), respectively. Vertical dashed red line indicates the BBH mean delay time.}
    \label{fig:LifeTimeClusters}
    \end{figure}

As shown in Fig. \ref{fig:LifeTimeClusters}, the cluster lifetimes are sensitive to their spatial distribution within the galaxy. In our reference model (black histogram), we assume that all GCs lie at an orbital radius of r$_g$ = r$_h$ (see section \ref{sec:Coupling}). To quantify the impact of this assumption, we performed a sensitivity test at z=6 by varying the orbital radii of $\pm$30\%. \\
In the r$_g$ = 0.7 r$_h$ case (red histogram), the increased frequency of encounters with molecular clouds in the denser inner regions leads to significantly shorter lifetimes, with the distribution peak shifting below 100 Myr. Conversely, at r$_g$ = 1.3 r$_h$ (green histogram), the reduced environmental density allows for much longer survival times, often exceeding 1 Gyr.\\
In Fig. \ref{fig:LifeTimeClusters}, we also show the average BBH delay time ($\sim$ 400 Myr, i.e., the time elapsed between the formation of the BBH and its merger). This value aligns almost perfectly with the peak of the lifetime distribution in our reference model. In contrast, the r$_g$ = 0.7 r$_h$ scenario would imply that a large fraction of GCs are disrupted before their internal BBH populations have sufficient time to merge, potentially underestimating the GW event rate. In the r$_g$ = 1.3 r$_h$ case, the extended cluster lifetimes would allow for a higher number of dynamical interactions leading to a higher BBH merger rate. These results indicate that while r$_g$ = r$_h$ works as a representative average environment.

\section{Impact of the gas surface density threshold on the globular cluster formation environments}
\label{app:GasSurfaceDensityAnalysis}

\begin{figure*}[h!]
     \centering
     \begin{subfigure}[b]{0.49\textwidth}
         \centering
         \includegraphics[width=\textwidth]{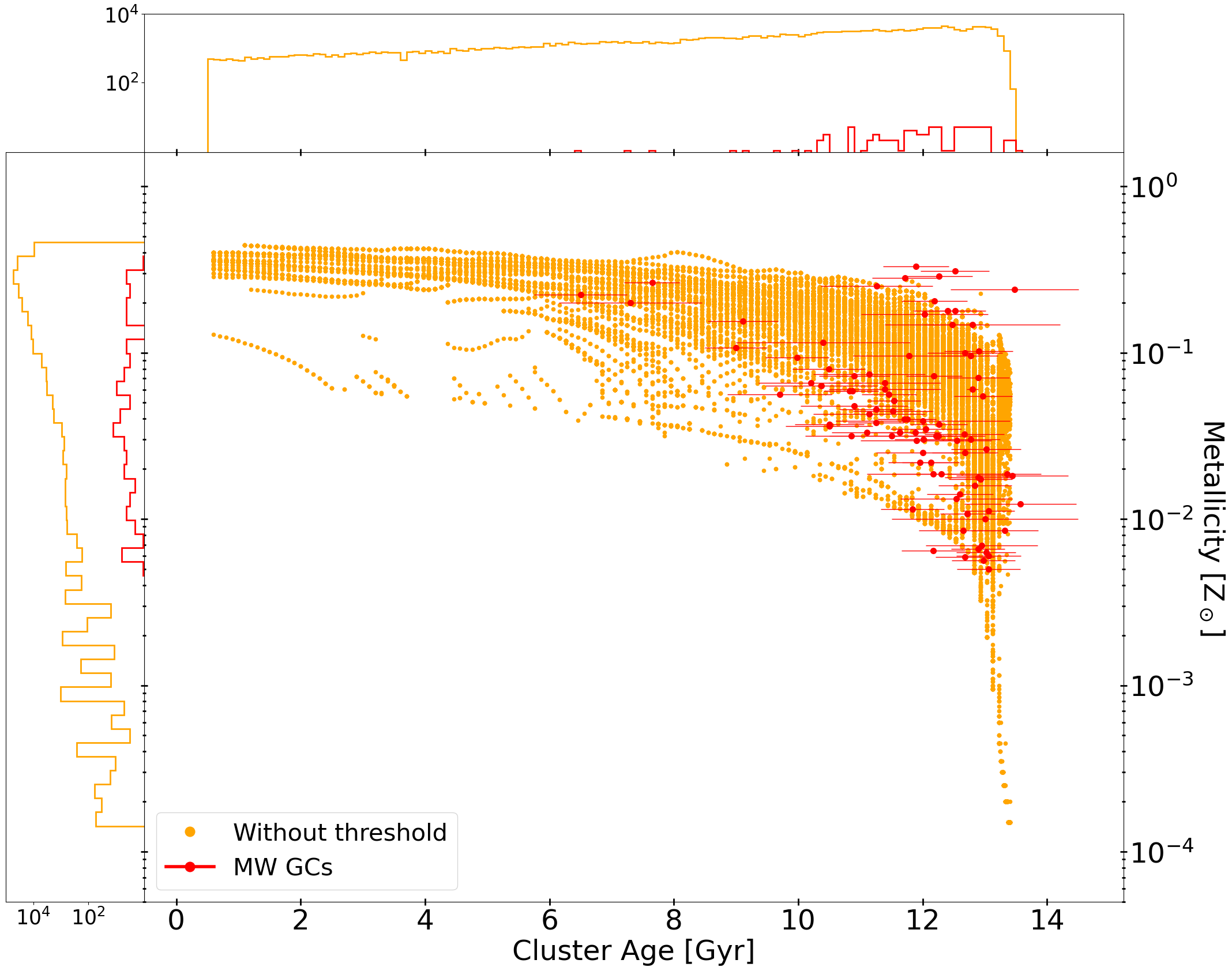}
         \caption{Without threshold}
         \label{fig:No_GSD}
     \end{subfigure}
     \hfill
     \begin{subfigure}[b]{0.49\textwidth}
         \centering
         \includegraphics[width=\textwidth]{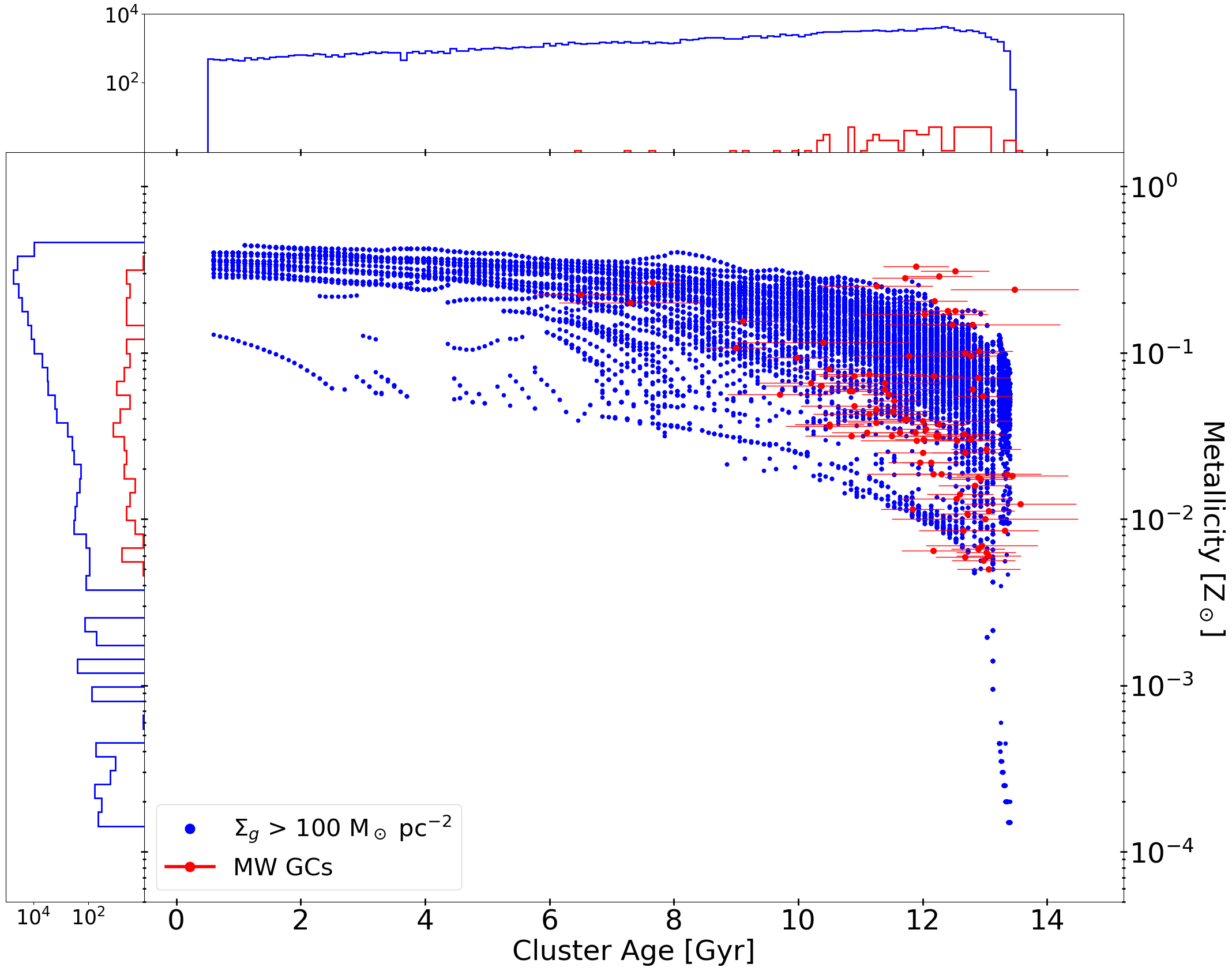}
         \caption{$\Sigma_g$ > 100 M$_\odot$ pc$^{-2}$}
         \label{fig:GSD100}
     \end{subfigure}

     \vspace{10pt} 
     \begin{subfigure}[b]{0.49\textwidth}
         \centering
         \includegraphics[width=\textwidth]{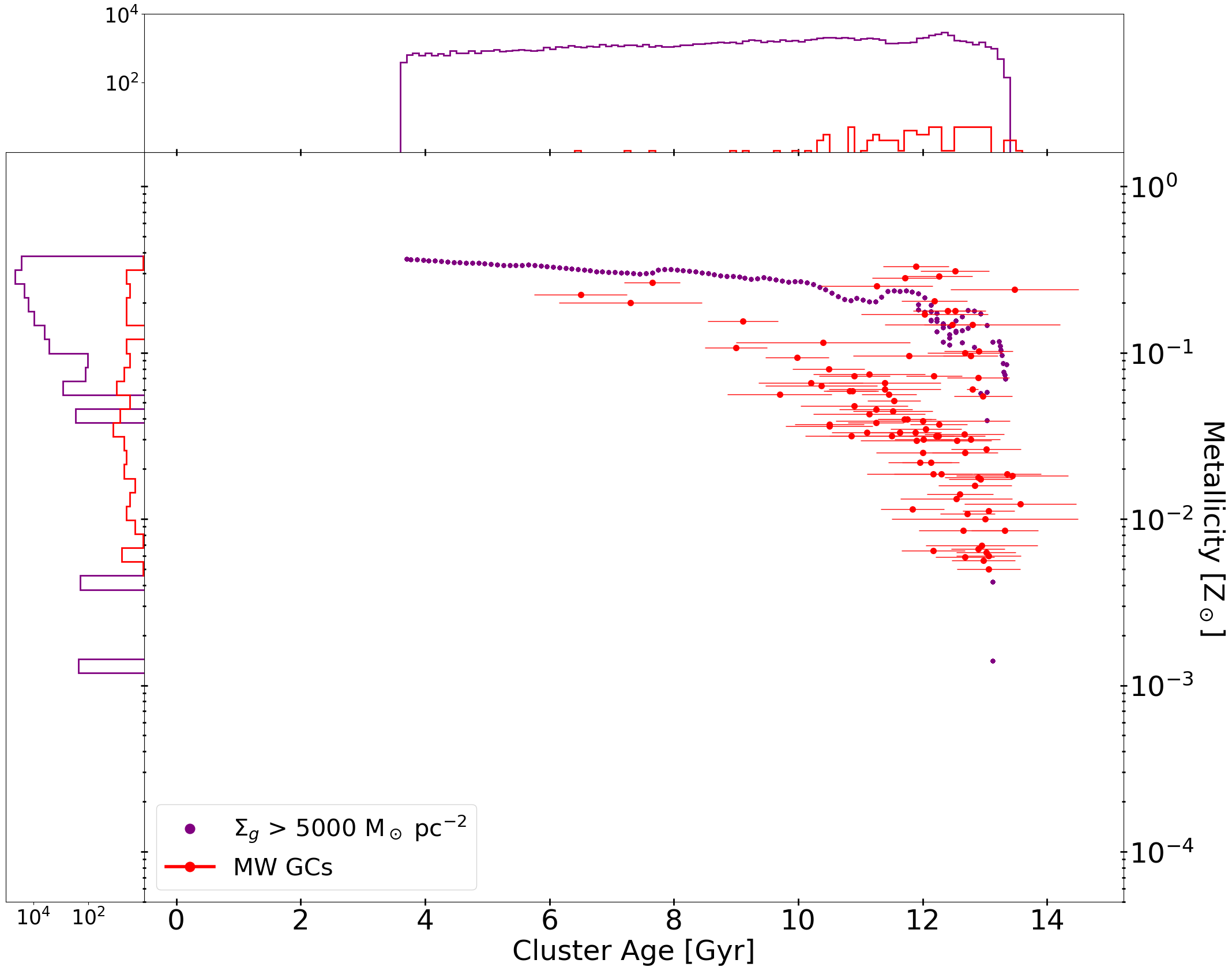}
         \caption{$\Sigma_g$ > 5000 M$_\odot$ pc$^{-2}$}
         \label{fig:GSD5000}
     \end{subfigure}
     \hfill
     \begin{subfigure}[b]{0.49\textwidth}
         \centering
         \includegraphics[width=\textwidth]{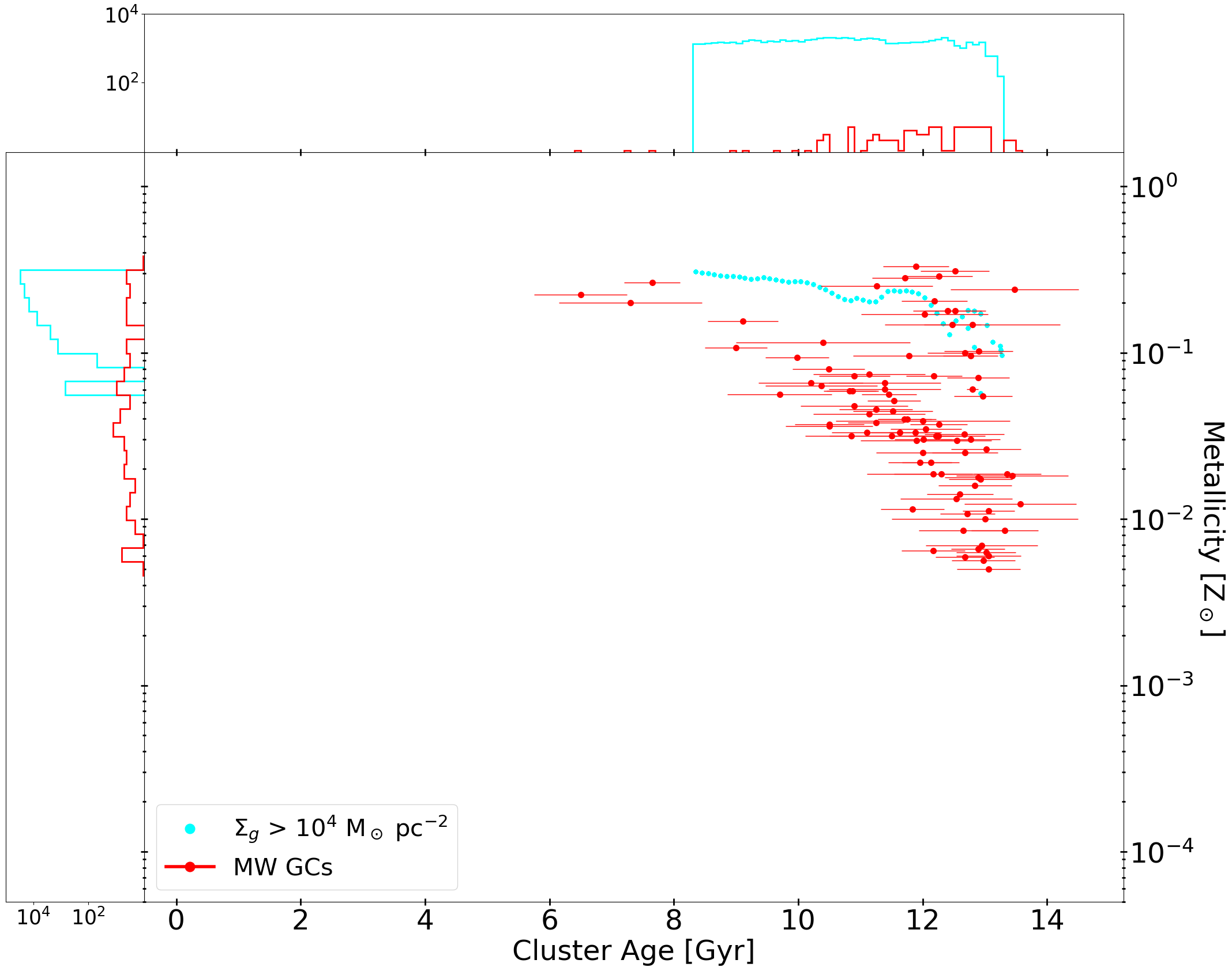}
         \caption{$\Sigma_g$ > $10^4$ M$_\odot$ pc$^{-2}$}
         \label{fig:GSD10000}
     \end{subfigure}

     \caption{Age-metallicity distribution of all potential GC formation environments within our LG-like volume, assuming different values for the $\Sigma_g$ threshold: (a) without threshold, (b) $\Sigma_g$ > 100 M$_\odot$ pc$^{-2}$, (c) $\Sigma_g$ > 5000 M$_\odot$ pc$^{-2}$, (d) $\Sigma_g$ > $10^4$ M$_\odot$ pc$^{-2}$). Red dots with error bars correspond to individual GCs observed in our Galaxy \citep{Kruijssen+2015}.}
     \label{fig:GSD_Analysis}
\end{figure*}
The gas surface density threshold is a critical parameter in distinguishing the galaxies of our LG-like volume capable to form GCs. As illustrated in Fig. \ref{app:GasSurfaceDensityAnalysis}, varying this threshold significantly impacts the number of host environments. We tested four values - above and below our main configuration - to evaluate the sensitivity of our results to this parameter. \\
If the threshold is high ($> 1000$ M$_\odot$ pc$^{-2}$), GC formation is restricted only to massive galaxies like the MW and Andromeda, failing to account for the spread of observational data. Indeed, we could not explain the metal-poor GCs. Conversely, an excessively low (or absent) threshold leads to overpredict the specific frequency of GCs in the LG, since also extremely metal-poor environments (Z $\approx$ 10$^{-4}$ Z$_\odot$) become GC hosting galaxies. These variations also inevitably influence the GC survival fraction within the MW halo. In the cases with $\Sigma_g > 5000$ M$_\odot$ pc$^{-2}$, GC formation is suppressed in almost all progenitor galaxies except for the main MW progenitor. Consequently, we found $\sim130$ surviving GCs in the MW halo at \textit{z} = 0 originate exclusively from the MW disk. This scenario fails to account for the significant population of ex situ surviving GCs known to be accreted from satellite galaxies. Conversely, the other two cases allow even small satellite galaxies to form GCs. This leads to an overproduction ($\sim300$ surviving GCs in the MW halo), as every minor and major merger event efficiently injects GCs into the halo.\\
Our reference model, with $\Sigma_g > 1000$ M$_\odot$ pc$^{-2}$, yields $185$ surviving GCs in the MW halo, providing the best agreement with both observed GCs and the expected contribution from accreted satellites. We therefore selected this value as our fiducial configuration.

\end{appendix}

\end{document}